\documentclass[reprint]{revtex4-2}
\usepackage{amsmath,amssymb,marginnote}
\usepackage{graphicx,subfigure}
\usepackage{dcolumn}
\usepackage{bm}
\usepackage[mathlines]{lineno}
\usepackage[colorlinks=true, allcolors=blue]{hyperref}
\usepackage[usenames]{color}
\newcommand{\liu}[1]{\textcolor{black}{#1}}

\begin{document}
\newcommand\barparent[1]{\overset{\scriptscriptstyle(-)}{#1}}
\title{Universality of the Neutrino Collisional Flavor Instability in Core Collapse Supernovae}
\author{Jiabao Liu}
\affiliation{Department of Physics and Applied Physics, School of Advanced Science \& Engineering, Waseda University, Tokyo 169-8555, Japan}
\author{Hiroki Nagakura}
\affiliation{Division of Science, National Astronomical Observatory of Japan, 2-21-1 Osawa, Mitaka, Tokyo 181-8588, Japan}
\author{Ryuichiro Akaho}
\affiliation{Department of Physics and Applied Physics, School of Advanced Science \& Engineering, Waseda University, Tokyo 169-8555, Japan}
\author{Akira Ito}
\affiliation{Department of Physics and Applied Physics, School of Advanced Science \& Engineering, Waseda University, Tokyo 169-8555, Japan}
\author{Masamichi Zaizen}
\affiliation{Faculty of Science and Engineering, Waseda University, Tokyo 169-8555, Japan}
\author{Shoichi Yamada}
\affiliation{Department of Physics, School of Advanced Science \& Engineering, Waseda University, Tokyo 169-8555, Japan}
\affiliation{Research Institute for Science and Engineering, Waseda University, Tokyo 169-8555, Japan}
\date{\today}
\begin{abstract}
     Neutrinos are known to undergo flavor conversion processes among the three flavors. The fast flavor conversion (FFC) has been the central piece of flavor conversions taking place in core-collapse supernovae (CCSNe) due to its shorter timescale to the completion of flavor conversion compared to other types of flavor conversion. Although the ordinary collisions between neutrinos and matter were once thought to decohere neutrinos and thus damp flavor conversions, it was recently realized that they can also induce the flavor conversion. The linear analysis showed that the so-called collisional flavor instability or CFI occurs in the absence of FFC. In this paper, we investigate if CFI takes place in of the post-bounce core of CCSNe, using the results of spherically symmetric Boltzmann simulations of CCSNe for four progenitor models with different masses. We also provide a\liu{n empirical correlation between} matter properties \liu{and} the occurrence of CFI in optically thick and semi-transparent regions; baryon mass density ($\rho$), electron fraction ($Y_e$), and the degeneracy of electron-type neutrinos ($\eta_{\nu_e}$) need to be $10^{10} {\rm g/cm^3} \lesssim \rho \lesssim 10^{12} {\rm g/cm^3}$, $Y_e\lesssim 0.4$, and $\eta_{\nu_e} \lesssim 0.5$, respectively. This condition allows us to
     easily locate the place of possible CFI occurence without detailed stability analyses, which is useful for analyzing CFI in CCSN models phenomenologically. 
\end{abstract}
\maketitle
\section{Introduction}\label{sec:Intro}
A star of mass $\gtrsim 10\,M_\odot$ undergoes a catastrophic gravitational collapse in the inner iron core when the matter density and temperature become high enough to trigger electron captures or photodisocciations of heavy nuclei, marking the onset of core-collapse supernova (CCSN). During the gravitational collapse and the early stage of post-bounce phases ($\lesssim 30$ ms after core bounce), electron-type neutrinos ($\nu_e$) are produced abundantly through charged-current processes. They are dominant carriers of energy and lepton number from the core to the outside of star. At $t \gtrsim 30$ ms, other species of neutrinos, electron-type anti-neutrinos ($\bar{\nu}_e$) and $\mu$- and $\tau$ neutrinos and anti-nneutrinos (collectively denoted as $\nu_x$), are also produced and escape from the hot and dense proto-neutron star (PNS) surface. The neutrino radiation from the CCSN core continues to cool down the PNS, dictating the thermal- and chemical evolutions towards the cold neutron star.

The neutrinos emitted from the PNS surface can undergo interaction with the intervening matter before escaping from the post-shock region. Some of $\nu_e$ and $\bar{\nu}_e$ are absorbed via charged-current reactions by nucleons, which transfers the neutrino energy to matter, a process known as neutrino reheating, which revitalizes the stagnated shock wave, being also enhanced by multi-dimensional fluid instabilities. According to recent detailed CCSN simulations, the shock revival has been observed rather commonly for a wide range of progenitor mass; see, e.g., (\cite{Ott_2018,10.1093/mnras/stz3223,10.1093/mnras/staa1048,Harada_2020,Bollig_2021,10.1093/mnras/stac1586,PhysRevD.107.043008}). Nowadays, there is emerging a consensus among different CCSN groups that the delayed neutrino heating mechanism accounts for a majority of CCSN explosions.

One thing we should mention, however, is that there remain many unresolved issues in both micro- and macroscopic physical processes even in up-to-date multi-dimensional models. One of the greatest uncertainties in any modeling of CCSNe is the neutrino quantum kinetics including flavor conversions. When neutrinos are dense, which is typically the case in the core region of CCSNe, neutrino self-interactions can induce collective flavor oscillations \cite{PANTALEONE1992128}. Since the self-interactions are intrinsically non-linear processes, the resultant flavor conversion has distinct features from those in vacuum and/or in matter. Fast neutrino-flavor conversion or FFC is an example \cite{PhysRevD.46.510,Sigl1993GeneralKD,doi:10.1146/annurev.nucl.012809.104524,CHAKRABORTY2016366,doi:10.1146/annurev-nucl-102920-050505,https://doi.org/10.48550/arxiv.2207.03561,PhysRevD.107.043024,PhysRevD.107.123024} and is currently attracting much attention. It can evolve with the timescale of picosecond in the CCSN core, which is much shorter than any relevant timescale of other physical processes, exhibiting the potential to greatly change the neutrino radiation field. It is also worth noting that multi-dimensional fluid instabilities facilitate the occurrence of FFC in the post-shock region. In fact, whereas angular crossings in the neutrino-flavor-lepton-number (NFLN), corresponding to the necessary and sufficient condition for FFC (see, e.g., \cite{PhysRevD.105.L101301}), are unlikely to appear in spherically symmetric CCSN models \cite{Tamborra_2017}, they have been observed rather commonly in multi-dimensional ones \cite{PhysRevD.100.043004,Nagakura_2019,PhysRevD.101.023018,PhysRevD.101.043016,PhysRevD.101.063001,Abbar_2020,PhysRevD.103.063013,PhysRevD.104.083025,Harada_2022,Akaho_2023,PhysRevD.99.103011,PhysRevD.101.023018,PhysRevD.101.043016,PhysRevD.100.043004,PhysRevD.104.083025,Harada_2022,Akaho_2023}. This has motivated the detailed studies of non-linear evolution of FFC \cite{PhysRevD.101.043009,PhysRevD.102.103017,PhysRevD.104.103003,PhysRevLett.128.121102,PhysRevD.104.103023,PhysRevD.106.103039,PhysRevLett.129.261101,PhysRevD.107.103022,PhysRevD.107.123021,xiong2023evaluating}, including their interplay with neutrino-matter collisions \cite{PhysRevD.103.063001,PhysRevD.105.043005,PhysRevD.106.043031,PhysRevD.103.063002,Kato_2021,10.1093/ptep/ptac082,PhysRevD.105.123003,Kato_2022,PhysRevD.106.103031,PhysRevD.103.063002,PhysRevLett.122.091101}. Some recent studies also demonstrate that neutrino radiation fields with FFC in the CCSN core are qualitatively different from those obtained from the classical neutrino transport \cite{PhysRevD.107.063025,PhysRevLett.130.211401,nagakura2023basic}, indicating that fluid dynamics, nucleosynthesis, and neutrino signals could be significantly impacted by FFC.

There is a caveat, however to the occurrence of FFC in the optically thick region, though. The angular distributions of $\nu_e$ and $\bar{\nu}_e$ are both almost isotropic in this region and NFLN angular crossings hardly occur unless the number densities of $\nu_e$ and $\bar{\nu}_e$ are very close to each other. According to the recent studies by \cite{PhysRevD.101.023018,PhysRevD.101.063001}, the convective region in PNS can offer a possibility, having almost the same $\nu_e$ and $\bar{\nu}_e$ number densities. However, these regions fluctuate in time violently with the dynamical timescale of PNS (see, e.g., \cite{10.1093/mnras/staa261}), and more importantly, they are usually very narrow, which may limit the impact of FFC on CCSN dynamics.

\citet{PhysRevLett.130.191001} recently pointed out that flavor conversions can be driven by the disparity in collision rates between different neutrino flavors and called it the collisional flavor instability or CFI. Thier properties in both linear- and non-linear regimes have been investigated from many different points of view \cite{PhysRevD.106.103031,PhysRevD.106.103029,duan,PhysRevD.107.083016,PhysRevD.106.103031,2022,PhysRevD.107.123011,2022,kato2023collisional}. There are mainly two noticeable features in CFI: (1) the timescale of CFI becomes shorter with increasing neutrino-matter interactions as well as the number density of neutrinos; (2) the instability can take place even if neutrino distributions are isotropic. These properties indicate a possibility that CFI occurs widely in the optically thick region, which is a clear advantage over FFC. In fact, the authors in \cite{PhysRevD.107.083016} investigated both the linear- and non-linear properties of CFI for \liu{a few} given CCSN fluid background \liu{snapshots}. They found that CFIs can occur ubiquitously (including optically thick regions), and the resultant neutrino radiation field is very different from those modeled by the classical neutrino transport.

On the other hand, the authors in \cite{shalgar2023neutrinos} also carried out recently similar simulations of the classical neutrino transport for a fluid profile taken from a CCSN model in \cite{mpaarchive}. Their results are not consistent with those in \cite{PhysRevD.107.083016}, however. The CFI regions in their simulations are much narrower than those in \cite{PhysRevD.107.083016}, and CFI is overwhelmed by FFC and even by slow mode (another type of collective neutrino oscillations; see, e.g., \cite{doi:10.1146/annurev.nucl.012809.104524}). There are many potential sources of the differences between the two works: for instance the neutrino radiation field is derived by the multi-energy-group treatment in \cite{PhysRevD.107.083016} while it is obtained with the gray approximation in \cite{shalgar2023neutrinos}; the \liu{hydrodynamical background} employed in \cite{PhysRevD.107.083016} is taken from a 18 $M_\odot$ CCSN model \liu{with muons} in \cite{1993ApJ...405..669M}, while \liu{that} in \cite{shalgar2023neutrinos} is \liu{extracted from a 18.6 $M_\odot$ CCSN model without muons}\liu{; while both works consider spherical symmetric CCSN models, \cite{shalgar2023neutrinos} in addition employs a mixing length scale scheme generating multi-dimensional convection effects favoring FFC.} As such, each work has its own pros and cons, hampering us from drawing robust conclusions about the occurrence of CFI in CCSN.

In this study, we attempt to settle the dispute by carrying out a systematic study on the occurrence of CFI with modern spherically symmetric CCSN models derived with the full Boltzmann neutrino transport. We pay an attention also to the growth rate of CFI if it is detected. We employ four different progenitor models, 11.2, 15, 27, and 40 $M_{\odot}$, and we explore the occurrence of CFI from core bounce to the late post-bounce phase ($> 400$ms) by sampling a matter profile every $\sim 1$ms. This study is the first comprehensive survey of the possible occurrence of CFI in CCSN, which has the ability to answer some intriguing questions: how common or rare is CFI?; when does CFI first appear in the post-bounce phase?; are there any progenitor-dependent features or universality?; does the unstable region appear persistently or intermittently? We also make an attempt to find correlation study between the occurrence of CFI with some matter properties, which potentially allows us to assess the possibility of CFI easily without conducting a detailed linear stability analysis. As shall be discussed below, however, the non-local effects of heavy-leptonic neutrinos tend to smear out the correlation, and consequently our analysis can provide only a\liu{n empirical} condition for the occurrence of CFI. It should be also mentioned that this work will be a stepping stone for our forthcoming work to explore CFI in multi-dimensional CCSN models (Akaho et al. in prep).

This paper is organized as follows. In Sec.~\ref{sec:CFIscheme}, we describe our method of linear stability analysis for CFI. After summarizing our CCSN models briefly in Sec.~\ref{sec:CCSNmodels}, we present our results in Sec.~\ref{sec:results}. Finally, we conclude the present study with future perspects in Sec.~\ref{sec:conclusion}. Throughout this paper, we use the metric signature of $+ - - -$. Unless otherwise stated, the natural unit $c = \hbar = 1$ is employed where $c$ and $\hbar$ are the light speed and the reduced Planck constant, respectively, when we discuss the governing equations for CFI (in Sec.~\ref{sec:CFIscheme}).

\section{Linear stability analysis of collisional flavor instability}\label{sec:CFIscheme}
The most straightforward way to assess the occurrence of CFI is to solve the dispersion relation, obtained by linearizing the quantum kinetic equation. In our previous paper \cite{PhysRevD.107.123011}, we developed a numerical scheme to solve the dispersion relation and provided analytic formulae to quantify growth rates of CFI approximately. These analytic formulae reduces computational costs and allows us to carry out a systematic survey of the occurrence of CFI in the post-bounce phase for multiple progenitor models. In this section, we summarize the essence of our method with the description of some approximations and assumptions to derive the analytic formulae. We also refer readers to our previous paper \cite{PhysRevD.107.123011} for more complete explanations of our stability analysis.

In this study, we work in the two flavor framework, which gives the same dispersion relation as that in the three flavor one, for the case with $\nu_{\mu}=\nu_{\tau}$. This is consistent with our CCSN models, in which all heavy leptonic neutrinos are assumed to be identical, and they are collectively denoted by $\nu_x$. We express the neutrino quantum kinetics in terms of the neutrino flavor density matrix,
\begin{equation}
    \rho(x,P)=
    \left(\begin{array}{cc}
        f_{\nu_e} & S_{ex} \\
        S_{xe}    & f_{\nu_x}
    \end{array}\right),
    \label{denmat}
\end{equation}
where the arguments $x$ and $P=(E,\mathbf{v})$ are the spacetime position and the 4-momentum of neutrinos, respectively; $E$ and $\mathbf{v}$ denote the neutrino energy and velocity, respectively, and the four velocity of neutrinos is $v \equiv (1, \mathbf{v})$ in the relativistic limit. For convenience, the flavor isospin convention is used hereafter, so that negative energy quantities are meant for antineutrinos: $\rho(E)=-\bar{\rho}(-E)$ for $E<0$. The time evolution of the neutrino flavor density matrix is governed by the quantum kinetic equation
\begin{equation}
    \mathrm{i}v\cdot\partial\rho=\lbrack H,\rho\rbrack+\mathrm{i}C,
    \label{fullqke}
\end{equation}
where $H$ represents the neutrino oscillation Hamiltonian and $C$ is the collision term. The Hamiltonian has the vacuum ($H_{\text{vac}}$), matter ($H_{\text{mat}}$), and neutrino self-interaction ($H_{\nu}$) contributions, which can be expressed as
\begin{equation}
\begin{split}
    &H=H_{\text{vac}}+H_{\text{mat}}+H_{\nu},\\
    &H_{\text{vac}}(x,P)=\frac{M^2}{2E},\\
    &H_{\text{mat}}(x,P)=\sqrt{2}G_\text{F} v\cdot \text{diag}(j_e(x),j_x(x)),\\
    &H_{\nu}(x,P)=\sqrt{2}G_\text{F}v\cdot\int\mathrm{d}P'\rho(x,P')v',\\
\end{split}
\end{equation}
where $M^2$ is the neutrino mass-squared matrix; $j_\alpha(x)$ is the lepton number 4-current of the charged lepton species $\alpha$; the integral over 4-momentum is abbreviated as
\begin{equation}
    \int dP=\int_{-\infty}^{\infty}\frac{E^2dE}{2\pi^2}\int\frac{d\boldsymbol{v}}{4\pi}.
\end{equation}
The collision term $C$ is written in the relaxation approximation as
\begin{equation}
    C(x,P)=\frac{1}{2}\{\text{diag}(\Gamma_{\nu_e}(x,P),\Gamma_{\nu_x}(x,P)),\rho_\text{eq}-\rho\}, \label{eq:colterm}
\end{equation}
where the curly bracket denotes the anti-commutator; $\Gamma_{\nu_\alpha}(x,P)$ is the collision rate for the neutrino of flavor $\alpha$; $\rho_\text{eq}$ denotes
the density matrix for the equilibrium state of the collisional processes. In this study, we take into account all emission and absorption interactions employed in our CCSN models (see Sec.~\ref{sec:CCSNmodels} for more details), but scattering processes are neglected. This assumption is in line with our approach deriving analytic formulae to estimate the growth rates of CFI (see below for more details). It should be mentioned that pair processes, e.g., electron-positron pair creation, nucleon-nucleon bremsstrahlung, and their inverse processes, can not be written in the form of Eq.~\ref{eq:colterm}, and the exact expression involves neutrino-momentum integrals (see \cite{PhysRevD.99.123014}). Handling them exactly is hence computationally expensive, so we utilize an approximate treatment here. Our CCSN models provide the total absorption neutrino opacities of each energy and species of neutrinos, in which the pair processes are also included. We estimate $\Gamma$ from them for each species of neutrinos, assuming that the collision term has the same form as Eq.~\ref{eq:colterm}. Although this prescription is pragmatic, it is a reasonable approximation not only in the optically thick but also in the semi-transparent regions where CFI would play an important role. We also note that the pair processes contribute to the collision term of $\nu_x$, indicating that $\Gamma_{\nu_x}$ is nonzero.

Assuming $|S_i| \ll f_{\nu_j}$, the off-diagonal element of Eq.~\ref{fullqke} can be linearized as
\begin{equation}
  \begin{split}
  & v\cdot(\mathrm{i}\partial-\Lambda_{0e}+\Lambda_{0x})S_{ex} \\
  + & (f_{\nu_e}-f_{\nu_x})\sqrt{2}G_\text{F}\int\mathrm{d}P'v\cdot v'S_{ex}(P') \\
  + & \frac{1}{2E}\sum_{\zeta=e,x}(M^2_{e\zeta}S_{\zeta x}-S_{e\zeta}M^2_{\zeta x})+\mathrm{i}\Gamma_{ex}S_{ex}=0,
\label{lineom}
  \end{split}
\end{equation}
where $\Lambda_{0z}=\sqrt{2}G_\text{F}[j_{z}(x)+\int\mathrm{d}Pf_{\nu_z}(x,P)v]$ and $\Gamma_{\nu_{ex}}(E)=\left[\Gamma_{\nu_e}(E)+\Gamma_{\nu_x}(E)\right]/2$. We then take a plane-wave ansatz as
\begin{equation}
S_{ex(xe)}(x,P)=S_{ex(xe)}(k,P)e^{\mathrm{i}k\cdot x}, \label{eq:Plane}
\end{equation}
where $k=(\omega,\boldsymbol{k})$ is the 4-wavevector. By inserting Eq.~\ref{eq:Plane} into Eq.~\ref{lineom}, the general form of dispersion relation can be obtained. However, it is computationally demanding to solve the dispersion relation, while we can obtain simple analytic formulae by approximating Eq.~\ref{lineom}.

First, we set $H_{\text{vac}}=H_{\text{mat}}=0$. This condition is in accordance with the purpose of this study that we quantify the growth rates of pure CFI mode. Next, we only focus on the so-called $\boldsymbol{k}=\boldsymbol{0}$ mode, which usually offers the maximum growth rate of CFI\,\cite{PhysRevD.107.123011}. Third, we apply the stability analysis to angular-integrated neutrino distributions in momentum space. This prescription is motivated by the fact that the anisotropy of neutrino distributions plays a subdominant role for CFI \cite{PhysRevD.107.123011}. This approximation is also in line with our treatment of the collision term, in which the scattering processes are not included. Since in and out scatterings are exactly cancelled if neutrinos angular distributions are isotropic, these processes can be safely ignored. Finally, we use a monochromatic assumption. As discussed in our previous paper \cite{PhysRevD.107.123011}, the growth rate of CFI for a non-monochromatic energy distribution is almost identical to the one for the monochromatic distribution, if we substitute the number density and their mean collision rates in the former for the counterparts in the latter.

Given these conditions, we can solve the dispersion relation analytically, and the solutions can be written as
\begin{equation}
\omega_{\pm}=-A-\mathrm{i}\gamma\pm\sqrt{A^2-\alpha^2+\mathrm{i}2G\alpha},
\label{monoew}
\end{equation}
for the isotropy-preserving modes and
\begin{equation}
\omega_{\pm}=\frac{A}{3}-\mathrm{i}\gamma\pm\sqrt{\left(\frac{A}{3}\right)^2-\alpha^2-\mathrm{i}\frac{2}{3}G\alpha},
\label{monoeisob}
\end{equation}
for the isotropy-breaking modes (see \cite{PhysRevD.107.123011} to derive these formulae). In the above equations, the following notations are introduced:
\begin{equation}
G=\frac{\mathfrak{g}+\bar{\mathfrak{g}}}{2},\ A=\frac{\mathfrak{g}-\bar{\mathfrak{g}}}{2},\ \gamma=\frac{\Gamma+\bar{\Gamma}}{2},\ \alpha=\frac{\Gamma-\bar{\Gamma}}{2},
\label{defGAgmal}
\end{equation}
where $\mathfrak{g}=\sqrt{2}G_\text{F}\left(n_{\nu_e}-n_{\nu_x}\right)$ and $\Gamma=(\Gamma_{\nu_e}+\Gamma_{\nu_x})/2$. The same applies to the barred quantities for antineutrinos. 
The number density of neutrinos and mean collision rates are computed by
\begin{equation}
\begin{split}
    n_{\nu_i}&=\int dP \hspace{0.5mm} f_{\nu_i}(P),\\
    \Gamma_{\nu_i}&=\langle\Gamma\rangle_{\nu_i}=\frac{1}{n_{\nu_i}}\int dP  \hspace{0.5mm} \Gamma_{\nu_{ex}}(P)f_{\nu_i}(P).
\end{split}
\end{equation}
In this paper, we use Eqs.~\ref{monoew}~and~\ref{monoeisob} to estimate the growth rate of CFI. We note that the maximum growth rate is usually given from the isotropy-preserving branch.

It is note-worthy that flavor conversions associated with neutrino-self interactions play important roles in the CCSN dynamics only if they overwhelm the collision rate. In other words, regions where the inequality $\mathfrak{g}_{\nu_i}\gg\Gamma_{\nu_i}$ 
is satisfied are of our interest. Assuming the inequality, Eqs.~\ref{monoew} and \ref{monoeisob} can be rewritten in a more concise form,
\liu{
\begin{equation}
\text{max\,\lbrack Im\,}\omega\rbrack=\begin{cases}
    -\gamma+\frac{|G\alpha|}{|A|},& \text{if }A^2\gg |G\alpha|,\\
    -\gamma+\sqrt{|G\alpha|},& \text{if }A^2\ll |G\alpha|,
\end{cases}
\label{wapprpre}
\end{equation}
}
for the isotropy-preserving branch and
\liu{
\begin{equation}
\text{max\,\lbrack Im\,}\omega\rbrack=\begin{cases}
    -\gamma+\frac{|G\alpha|}{|A|},& \text{if }A^2\gg |G\alpha|,\\
    -\gamma+\frac{\sqrt{|G\alpha|}}{\sqrt{3}},& \text{if }A^2\ll |G\alpha|,
\end{cases}
\label{wapprbr}
\end{equation}
}
for the isotropy-breaking branch. It should be also mentioned that althoug the obtained growth rates are not exactly the same as those obtained by directly solving dispersion relation, we confirm that the error is within a factor around unity even in extreme cases. These concise expression also helps us to see if the resonance-like CFI occurs in our models \cite{PhysRevD.107.123011,2022}, which will be discussed in Sec.~\ref{subsec:reso}.

\section{CCSN models}\label{sec:CCSNmodels}
As described in Sec.~\ref{sec:CFIscheme}, CFI hinges on not only neutrino distributions but also collision rates, suggesting that an accurate radiation-hydrodynamic modeling of CCSN is required to make a robust and reliable analysis of CFI. We utilize our up-to-date CCSN models, in which all necessary data for our stability analysis, matter and neutrino distributions and collision rates, are provided. Below, we briefly summarize our CCSN models.

Details on our CCSN code can be found in \cite{Nagakura_2014,Nagakura_2017,Nagakura_2019c}. Although it has the ability to perform multi-dimensional simulations (see a series of our previous papers: \cite{Nagakura_2018,Harada_2019,Nagakura_2019acceleration,Harada_2020,Iwakami_2020,Iwakami_2022,Akaho_2023}), in this study we employ only its spherically symmetric capability. The code obtains the time evolution of neutrino radiation field by solving the Boltzmann equation under multi-energy, multi-angle, and multi-species ($\nu_e$, $\bar{\nu}_e$, and $\nu_x$) treatments. Hydrodynamics with Newtonian self-gravity is simultaneously solved with the self-consistent feedback from neutrino transport and its matter-interactions. We employ Togashi-Furusawa EOS based on a variational method, first developed by Togashi et al.\,\cite{TOGASHI201353,TOGASHI201778} and later extended to include an ensemble of nuclei by Furusawa et al.\,\cite{Furusawa_2017}. Various neutrino-matter interactions are incorporated in these simulations; electron-positron pair annihilation, nucleon-nucleon bremsstrahlung, and electron and positron captures by nucleons, heavy and light nuclei as emission processes; their inverse reactions as absorption ones; nucleon scatterings, electron-scatterings, and coherent scatterings with heavy nuclei as scattering processes. The detailed numerical implementations and approximations for these neutrino-matter interactions can be found in \cite{Nagakura_2019a,Sumiyoshi_2012}.

In \cite{Nagakura_2019a}, we ran CCSN simulations for four different progenitor models in \cite{RevModPhys.74.1015} with zero age main sequence progenitor masses of 11.2, 15, 27, and 40 $M_{\odot}$. Although all these fail to explode, both matter and neutrino dynamics are substantially different among models. For instance, the shock radius in the early post-bounce phase is remarkably larger in the 11.2~$M_{\odot}$ model than other models, while it has the lowest neutrino luminosity and PNS mass. We also note that the 40 $M_{\odot}$ model undergoes the highest mass accretion rate onto PNS, leading to the highest mass of PNS among models (which seems to eventually create a black hole even in multi-dimensional models; see, e.g., \cite{Ott_2018,burrows2023blackhole}). As such, these four models give some representative progenitor dependence of CCSN, and we will delve into whether they have an influence on the occurrence of CFI in this study.

\section{Results}\label{sec:results}
\subsection{Overall properties}\label{subsec:overP}

\begin{figure*}
    \centering
    \subfigure[]{\includegraphics[width=0.49\textwidth]{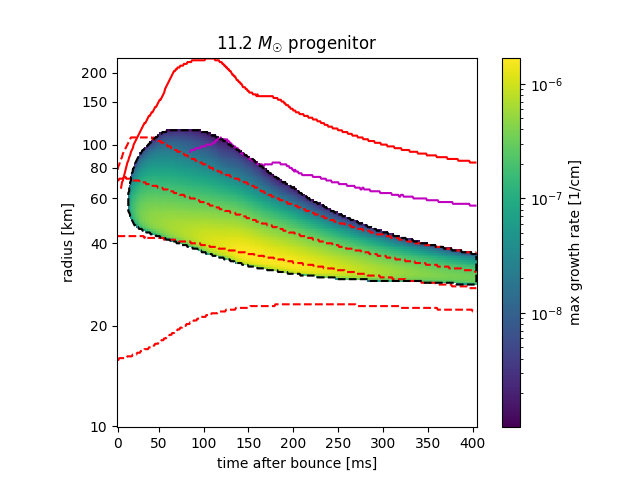}
         \label{rate1}
     }
     \subfigure[]{\includegraphics[width=0.49\textwidth]{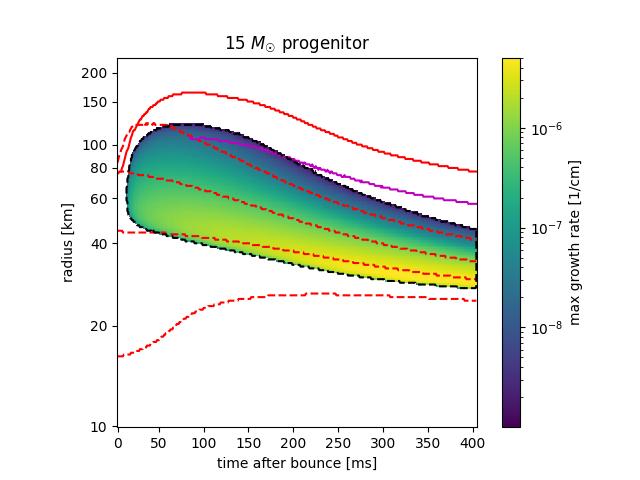}
         \label{rate2}
     }
     \subfigure[]{\includegraphics[width=0.49\textwidth]{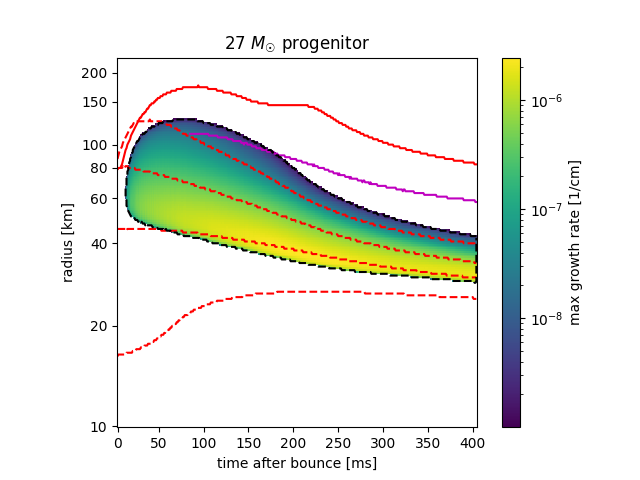}
         \label{rate3}
     }
     \subfigure[]{\includegraphics[width=0.49\textwidth]{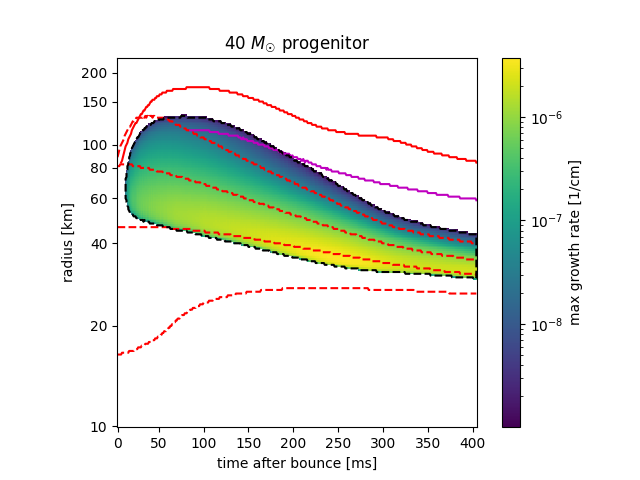}
         \label{rate4}
     }
     \caption{
     Radius-time diagram for the maximum growth rate of CFI ($\mathrm{max}\left[\mathrm{Im\,\omega}\right]$) estimated from Eq.\ref{monoew} and \ref{monoeisob} for progenitors of 11.2 (panel a), 15 (panel b), 27 (panel c), and 40 $M_\odot$ (panel d). In each panel, red and purple solid lines portray time trajectories of shock- and gain radius, respectively. The red dashed lines traces the isodensity radii of mass density: $10^{13},\ 10^{12},\ 10^{11},\ \text{and}\ 10^{10}\,\mathrm{g\,cm^{-3}}$.
     }
    \label{rateplot}
\end{figure*}

\begin{figure*}
    \centering
     \subfigure[]{\includegraphics[width=0.49\textwidth]{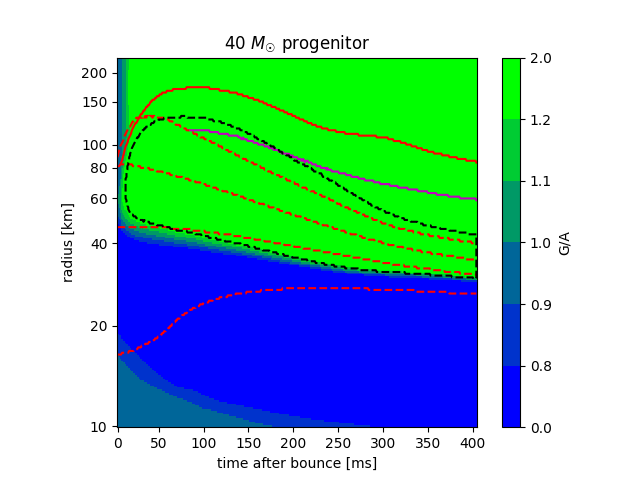}
         \label{GbyA}
     }    
     \subfigure[]{\includegraphics[width=0.49\textwidth]{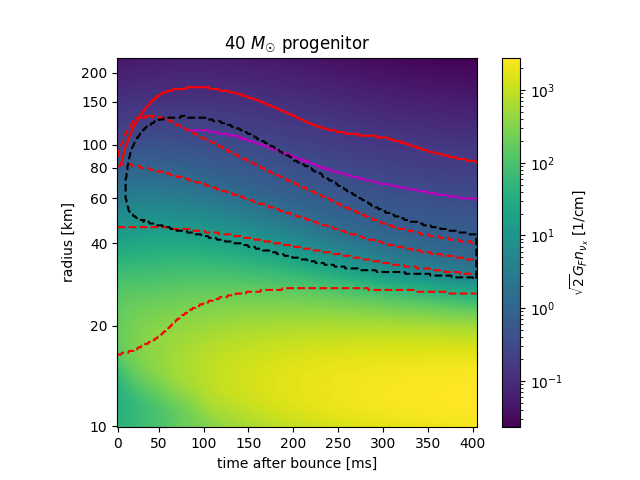}
         \label{gx}
     }
     \caption{
     Same as Fig.~\ref{rateplot} but for $G/A$ (panel a) and $\sqrt{2} G_{\mathrm{F}} n_{\nu_x}$ (panel b). In this figure, we only focus on results of $40 M_{\odot}$ model. The dashed black line in each figure corresponds to the boundary of CFI.
     }
    \label{numu1}
\end{figure*}

\begin{figure*}
    \centering
     \subfigure[]{\includegraphics[width=0.49\textwidth]{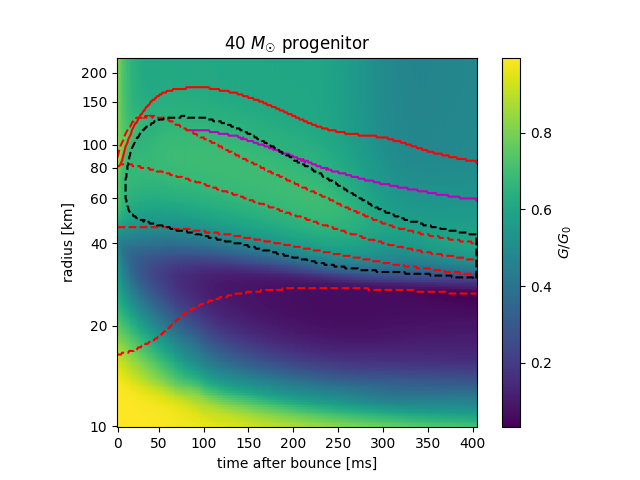}
         \label{gratio}
     }
    \subfigure[]{\includegraphics[width=0.49\textwidth]{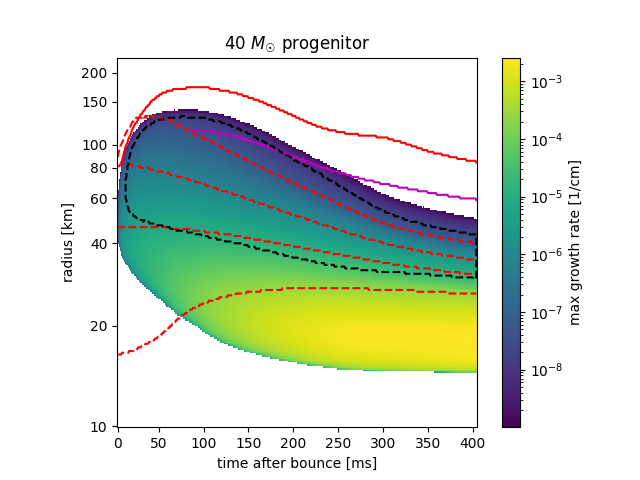}
         \label{ratermx}
     }
     \caption{
     Same as Fig.~\ref{numu1} but for $G/G_0$ (panel a) and maximum growth rate of CFI estimated from Eq.\ref{monoew} and \ref{monoeisob} with $G_0$ ($n_{\nu_x}$ is assumed to be zero); see the text for more details.
     }     
    \label{numu2}
\end{figure*}

\begin{figure*}
    \centering
    \subfigure[]{\includegraphics[width=0.49\textwidth]{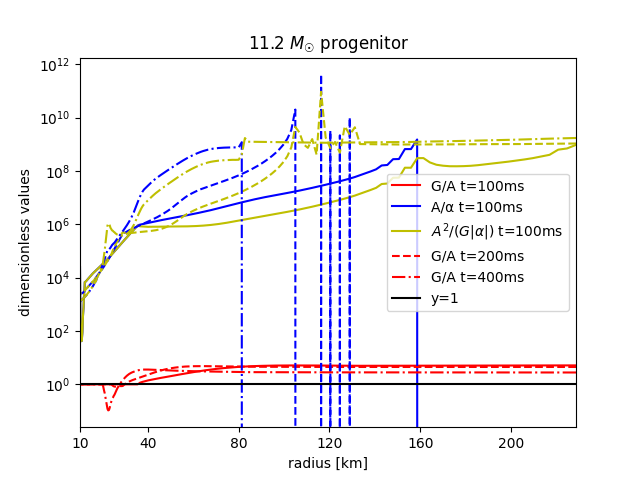}
         \label{line1}
     }
     \subfigure[]{\includegraphics[width=0.49\textwidth]{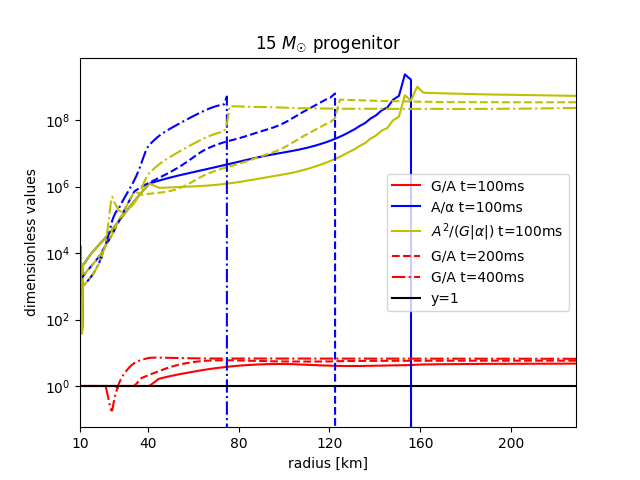}
         \label{line2}
     }
     \subfigure[]{\includegraphics[width=0.49\textwidth]{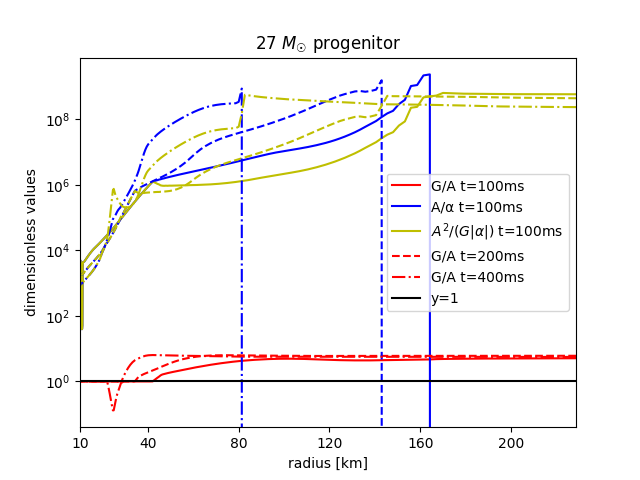}
         \label{line3}
     }
     \subfigure[]{\includegraphics[width=0.49\textwidth]{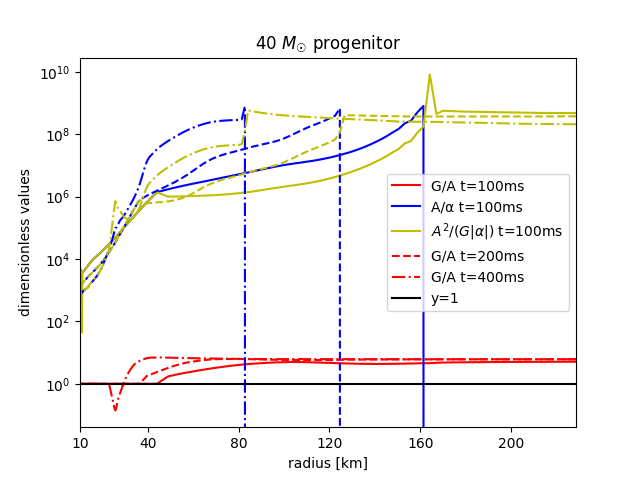}
         \label{line4}
     }
     \caption{
     Radial profiles of $G/A$ (red), $A/\alpha$ (blue), and $A^2/(G|\alpha|)$ (gold) at times $t=100,\ 200,\ 400\,\mathrm{ms}$ after bounce, while the line type denotes the time. Each panel distinguishes progenitor model.
     }
    \label{lineplots}
\end{figure*}

Figure~\ref{rateplot} displays the two-dimensional color map of maximum growth rate of CFI as functions of time (measured from core bounce) and radius, while different panels distinguish CCSN models. To guide the eye, we show the shock radius by the red solid line, gain radius by the purple solid line, isodensity radii of $\rho=10^{10},\ 10^{11},\ 10^{12},$ and $10^{13}\,\mathrm{g\,cm^{-3}}$ ($\rho$ denotes the mass density of baryons) as the dashed red lines. As another remark, we find that CFI can become unstable in the pre-shock regions. However, the growth rate is too small $\lesssim 10^{-9}\,\mathrm{cm^{-1}}$ to give any impacts on CCSN dynamics. For this reason, we set a minimum value of the growth rate at $10^{-9}\,\mathrm{cm^{-1}}$ in these plots.

As shown in this figure, CFI is commonly detected in the region of $10^{10}\,\mathrm{g\,cm^{-3}} \lesssim \rho \lesssim 10^{12}\,\mathrm{g\,cm^{-3}}$ and they can exist rather stably regardless of progenitors, suggesting that CFI will take place universally in post-shock regions. We also find that CFI can occur from the early phase ($\sim 20\,\mathrm{ms}$) and subsequently expands its region with time for both in and outward radial directions. In the late phase, however, the CFI region with $>10^{-9}\,\mathrm{cm^{-1}}$ shrinks with time. One might think that this is due to the failure of shock revival in our CCSN models, i.e., artifacts due to spherical symmetry. It should be noted, however, that CFI is dictated by neutrino-matter interactions in addition to the neutrino distributions and hence the time evolution of the CFI region is associated with matter distributions. As observed also in multi-dimensional successful explosion models, the CCSN core shinks with time, to form a neutron star, suggesting that the contraction of the CFI region in the late phase is mainly a consequence of this shrinkage of the core and will be generic even in multi-dimensional models.

We also find that CFI is not detected in very optically thick regions ($\rho \gtrsim 10^{13}\,\mathrm{g\,cm^{-3}}$). The absence of CFI inside the PNS ($r \lesssim 15$ km) is rather obvious, since $\nu_e$ is highly degenerate, leading to an extreme disparity of number densities between $\nu_e$ and $\bar{\nu}_e$. In addition to this, $\nu_x$ is also much less populated than $\nu_e$ due to low matter temperatures inside PNS. This implies that $A$ becomes nearly equal to $G$, resulting in the suppression of CFI (see Eqs.~\ref{wapprpre}~and~\ref{wapprbr}). On the other hand, the matter temperature is higher in the regions of $r \gtrsim 15$ km (since they experienced shock heating), indicating that both $\bar{\nu}_e$ and $\nu_x$ can be populated. As shown below, $\nu_x$ plays an important role on suppressing CFI in the region of $\rho \gtrsim 10^{13}\,\mathrm{g\,cm^{-3}}$.

In the following discussion, we focus on $40 M_{\odot}$ model, since we confirm that other progenitor models have the same trend. Fig.~\ref{GbyA} displays a 2D color map for the ratio of $G$ to $A$. One noticeable feature in this figure is that $|G/A|$ can be smaller than unity in the inner region (colored blue). This is a clear indication that $\nu_x$ becomes abundant. We note that $|G/A|$ should be higher than unity if there are no $\nu_x$ and their anti-partners (see Eq.~\ref{defGAgmal}). We also note that $G$ becomes smaller when $\nu_x$ appears, whereas $A$ remains constant under the condition of $\nu_x = \bar{\nu}_x$. As shown in Fig.~\ref{gx}, $|G/A|$ distributions have a clear correlation with the $n_{\nu_x}$ distribution.

Let us corroborate our claim that CFI is suppressed by $\nu_x$ in the optically thick region. In Fig.~\ref{gratio}, we portray a color map of $G/G_{0}$, where $G_{0}$ represents $G$ with $n_{\nu_x}$ assumed to be zero. The deviation of $G/G_0$ from unity hence corresponds to the contribution of $\nu_x$ to $G$. As clearly seen in the figure, the ratio in the region at $\rho \gtrsim 10^{13}\,\mathrm{g\,cm^{-3}}$ with $r \gtrsim 15$ km is remarkably smaller than unity, indicating that $n_{\nu_x}$ plays an important role. To strengthen this discussion, we carry out the same stability analysis of CFI by replacing $G$ with $G_{0}$. The result of the growth rate is shown in Fig.~\ref{ratermx}. As shown clearly in the figure, the inner boundary of the CFI region is located at much smaller radii than the case with $\nu_x \neq 0$ (see panel d in Fig.~\ref{rateplot}). This provides strong evidence that $\nu_x$ hampers an occurrence of CFI in the optically thick region.

\subsection{Resonance-like CFI}\label{subsec:reso}

\begin{figure}
    \centering
     \subfigure[]{\includegraphics[width=0.49\textwidth]{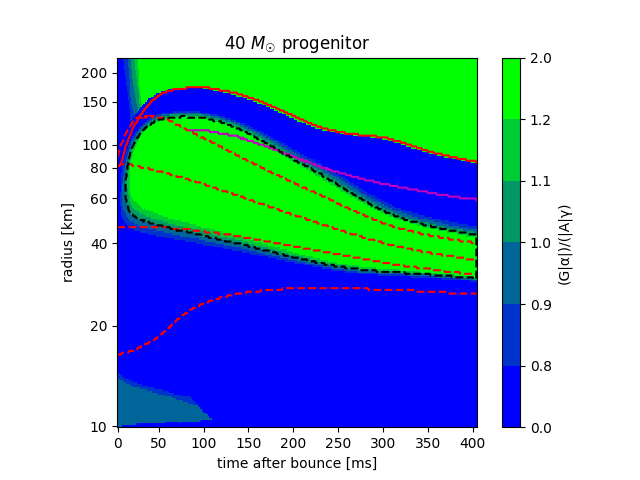}
         \label{4ratio4}
     }     
     \caption{
     Same as Fig.~\ref{numu1} but for $(G|\alpha|)/(|A|\gamma)$. We display the result of $40 M_{\odot}$ model as a representative case.
     }
    \label{fig:apGro}
\end{figure}

One of the unique properties of CFI is a resonance-like feature, in which the growth rate can be remarkably higher than the typical non-resonance value. If the resonance-like CFI occurs, the complete flavor swaps between $\nu_e$ and $\nu_x$, may take place \cite{kato2023collisional}, which potentially leads to a radical change in the neutrino radiation fields.

We do not find the resonance-like CFI in our CCSN models, however. We reached this conclusion by the following analysis. Before going into details, we briefly summarize the property of resonance-like CFI from Eqs.~\ref{wapprpre}~and~\ref{wapprbr}. The growth rate of CFI is comparable to the collision rate when $A^2\ll G|\alpha|$. In the region, where the neutrino self-interaction potential is larger than the collision rate, the condition $A^2\gg G|\alpha|$ is usually satisfied. However, if the number densities of $\nu_e$ and $\bar{\nu}_e$ approach each other, $A$ becomes lower and $A^2\ll G|\alpha|$ may be realized. Then the growth rate is proportional to $\sqrt{G \alpha}$.
Since $G$ is much larger than $\alpha$, the growth rate of CFI can be significantly larger, which accounts for the resonance-like feature.

In Fig.~\ref{lineplots}, we show the radial profiles of $A^2/G|\alpha|$ and its associated quantities for all progenitor models. We find that $A^2/G|\alpha|$ is much greater than unity in the entire post-bounce region, suggesting that the resonance condition is hardly achieved. It is interesting to note that $G/A$ can be larger than unity in the optically thick region, which facilitates the occurrence of the resonance-like CFI. As shown in Fig.~\ref{lineplots}, however, $A/|\alpha|$ is significantly higher than $G/A$, leading to $A^2/G|\alpha| \gg 1$. This analysis suggests that $A$ needs to be at least $\sim \alpha$ for the resonance-like CFI. On the other hand, the self-interaction potential is usually several orders of magnitude higher than the collision rate, implying that $A$ needs to be essentially zero for the resonance-like CFI, which is not realized in our CCSN models.

To strengthen our discussion, we also make a plot for $G |\alpha|/|A| \gamma$ for the $40 M_{\odot}$ model in Fig.~\ref{fig:apGro}. Note that the isotropy-preserving branch gives the maximum growth rate for the CFI out of resonance. This ratio of $G|\alpha|/(|A|\gamma)$ is associated with the growth rate of non-resonance CFI (see Eq.~\ref{wapprpre}) and needs to be greater than unity for the occurrence of CFI. As clearly shown in this figure, the region of $G |\alpha|/|A| \gamma > 1$ matches exactly that of CFI, indicating that CFI is not resonance-like. 

An important remark must be made here. No detection of the resonance-like CFI in our models may be an artifact of spherical symmetry. As shown in multi-dimensional CCSN models, PNS convections accelerate deleptonization of the inner core \cite{10.1093/mnras/staa261} and reduces the degeneracy of $\nu_e$ and leads to $n_{\nu_e} \sim n_{\bar{\nu}_e}$, i.e., $A \sim 0$. On the other hand, $\nu_x$ seems to be populated also in this region, implying that $G$ becomes also lower (see Sec.~\ref{subsec:overP}) and the growth rate of the resonance-like CFI, if any, will be suppressed. These considerations indicate a need of detailed and quantitative analyses of CFI based on realistic multi-dimensional CCSN models. We defer this intriguing study to a future work.

\subsection{Correlation between matter properties and CFI}\label{subsec:corMatCFI}

\begin{figure*}
    \centering
    \subfigure[]{\includegraphics[width=0.49\textwidth]{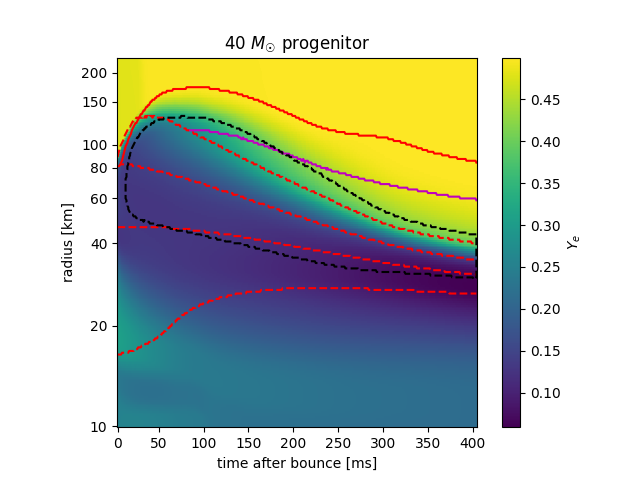}
         \label{Ye}
     }
     \subfigure[]{\includegraphics[width=0.49\textwidth]{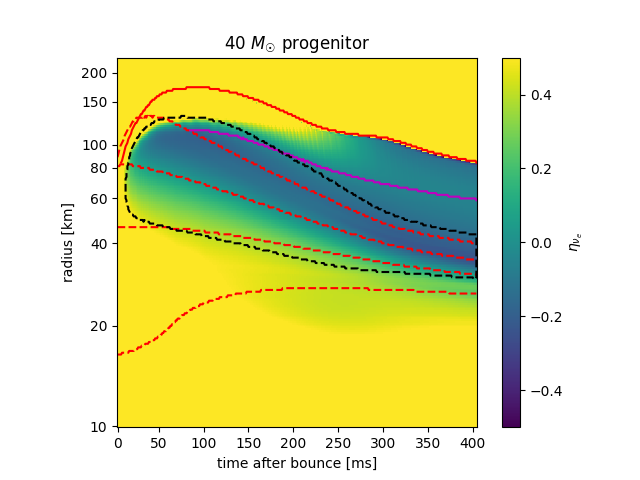}
         \label{dege}
     }
     \caption{
     Same as Fig.~\ref{fig:apGro} but for $Y_e$ (panel a) and $\eta_{\nu_e}$ (panel b).
     } 
    \label{hydroparam}
\end{figure*}

Our stability analysis suggests that CFI commonly occurs in the post-shock region of CCSN. The CFI region straddles the transition layer between the optically thick and the semi-transparent regions, where neutrinos and matter are interacting with each other frequently. One then expects that the CFI region has correlations with some local matter properties. The investigation of such correlations is highly beneficial to those assessing CFI in the phenomenological CCSN models, in which the neutrino radiation field may not be well modeled. The result can also be used to narrow down searching regions before performing a detailed stability analysis, which will reduce the computational cost of the search, in particular for multi-dimensional CCSN models.

As mentioned already, the matter density seems to be a good indicator; CFI tends to occur in $10^{10}\,\mathrm{g\,cm^{-3}} \lesssim \rho \lesssim 10^{12}\,\mathrm{g\,cm^{-3}}$. We also explore possible correlations of CFI with two other thermodynamic quantities: the electron fraction ($Y_e$) and the electron neutrino degeneracy ($\eta_{\nu_e}=\mu_{\nu_e}/T$), where $\mu_{\nu_e}$ denotes the chemical potential of $\nu_e$, which is defined as $\mu_p - \mu_n + \mu_{e}$ ($p, n,$ and $e$ denote free protons, neutrons, and electrons, respectively). We selected these quantities because they are the quantities to characterize $\nu_e$ and $\bar{\nu}_e$ in the optically thick region. Previous works showed positive correlations between the two quantities and FFC (see, e.g., \cite{PhysRevD.101.023018,PhysRevD.104.083025}).

In Fig.~\ref{hydroparam}, we show the $Y_e$ and $\eta_{\nu_e}$ distributions for the $40 M_{\odot}$ model. Although positive correlations can be seen, it seems hard to assess the occurrence of CFI only with these quantities. For instance, although we find CFI in the regions of $Y_e \lesssim 0.4$, stable regions are also existent there. A similar trend is also found in $\mu_{\nu_e}$: CFI can occur in the regions of $|\mu_{\nu_e}| \lesssim 0.5$, but not always.

Much of the complexity comes from $\nu_x$. As is well known, the $\nu_x$ distributions do not affect the occurrence of FFC (as long as $\nu_x=\bar{\nu}_x$), but they affect the occurrence of CFI (see Sec.~\ref{subsec:overP}). We also note that $\nu_x$ does not have charged-current reactions with matter and their energy sphere is located at smaller radii than those of $\nu_e$ and $\bar{\nu}_e$. This indicates that the $\nu_x$ number density can not be determined by the local equilibrium condition. This consideration suggests that the local matter property may not provide a sufficient condition. Nevertheless, our result suggests that conditions of $10^{10}\,\mathrm{g\,cm^{-3}} \lesssim \rho \lesssim 10^{12}\,\mathrm{g\,cm^{-3}}$, $Y_e \lesssim 0.4$, and $|\mu_{\nu_e}| \lesssim 0.5$ need to be satisfied for the occurrence of CFI in the optically thick and semi-transparent regions, which is useful to locate possible CFI regions. As an important remark, one needs to keep in mind that this condition could be altered in multi-dimensional cases, though.

\subsection{Comparison with previous studies}\label{subsec:compare}
As shown above, we find that CFI can occur rather commonly regardless of progenitors at $10^{10}\,\mathrm{g\,cm^{-3}} \lesssim \rho \lesssim 10^{12}\,\mathrm{g\,cm^{-3}}$ and forms a durable layer. This location corresponds to the transition layer for $\nu_e$ and $\bar{\nu}_e$ from the optically thick to the semi-transparent regions. This result supports the claim in \cite{PhysRevD.107.083016} that CFI can occur in the post-shock regions. The obtained growth rate is also qualitatively consistent with theirs.

On the other hand, resonance-like CFIs are not detected in our CCSN models (although this may be artifacts of spherical symmetry), implying that the growth rate of CFI is of the same order of magnitude as the collision rate. This also means that CFI is overwhelmed by FFC, if NFLN crossings appear, which is in line with suggestions by \cite{nagakura2023basic,shalgar2023neutrinos}. We note in addition that the growth rate of slow mode could be higher than that of CFI as suggested by \cite{shalgar2023neutrinos}. However, the slow modes seem to be hampered by high mass densities in the post-shock region during the accretion phase \cite{PhysRevD.78.085012,PhysRevD.85.065008,PhysRevLett.107.151101,PhysRevD.85.113007}. Some (unknown) mechanisms are required then to trigger the slow instability.

\section{Conclusion}\label{sec:conclusion}
In this paper, we carry out a systematic study on the occurrence of CFI based on the spherically symmetric CCSN models developed with the modern input physics and full Boltzmann neutrino transport. We use the approximate analytic formulae obtained in our previous study \cite{PhysRevD.107.123011} to quantify the growth rate of CFI. They are reasonable accurate and also much more computationally efficient than the numerical solution of the dispersion relation. We find that CFI occurs universally in the post-shock region regardless of progenitors.

The CFI region is roughly bounded by isodensity lines of $10^{12}\ \text{and}\ 10^{10}\,\mathrm{g\,cm^{-3}}$. It continues to exist rather stably at $\gtrsim 20 {\rm ms}$ after bounce. We also find that $\nu_x$ plays an important role to hamper CFI in the optically thick region, which is a feature of CFI that distinguishes it from FFC. This implies the importance of accurate modeling of the $\nu_x$ distribution, and $\nu_x$ should be taken into account appropriately in the CFI analyses.

It is worth noting that our results do not show any signs of resonance-like CFI. As a consequence the growth rate of CFI is of the same order as the collision rate. However, this result seems to be artifacts of spherical symmetry we imposed and may be qualitatively changed in multi-dimensional models. This is because convection in PNS layers enhances deleptonization there and will facilitate the occurrence of resonance-like CFI as well as FFC. On the other hand, it has been also suggested that the PNS convection will amplify the $\nu_x$ diffusion and may suppress CFI. These competing effects to CFI inherent in the multi-dimensional models should be investigated in detail with sophisticated CCSN models, which is the top priority in our future works.

The present study also reveals a difficulty of assessing CFI only by matter properties. We find that $\nu_x$ is mainly responsible for the complexity. It is attributed to the fact that the $\nu_x$ diffusion is greater than that of $\nu_e$ and $\bar{\nu}_e$ due to the lack of charged-current reactions. However, we find that CFI tends to occur in regions of $10^{10}\,\mathrm{g\,cm^{-3}} \lesssim \rho \lesssim 10^{12}\,\mathrm{g\,cm^{-3}}$, $Y_e \lesssim 0.4$, and $\eta_{\nu_e} \lesssim 0.5$, suggesting that these conditions are useful to narrow down the possible location of CFI.

Our systematic study of CFI provides independent and robust information, useful to settle the dispute in the previous works on CFI in the CCSN environments (see \cite{PhysRevD.107.083016,shalgar2023neutrinos}). Our study supports the results of \cite{PhysRevD.107.083016} that CFI occurs in a wide spatial range across the optically thick and semi-transparent regions. On the other hand, no occurrences of resonance-like CFI suggests that CFI would be overwhelmed by FFC or even by the slow instability (although the latter seems unlikely to be developed) if they occur \cite{shalgar2023neutrinos}. This is also consistent with our previous study \cite{nagakura2023basic}, in which the flavor conversions in the non-linear phase are mainly characterized by FFCs. It should be stressed again, however, that these results may be altered in multidimensions, \liu{where FFC will likely take place}. We, hence, should postpone the final conclusion on the impacts of CFI on CCSN dynamics until we can carry out detailed study of CFI and other flavor conversions based on multi-dimensional models.

There remain other intriguing open questions about CFI. As has been reported recently, muonization may happen notably in CCSNe\,\cite{10.1093/ptep/ptac118,PhysRevLett.119.242702,PhysRevD.102.023037,PhysRevD.102.123001}. The existence of muons changes the playground of muon and anti-muon neutrinos significantly and induces a number density disparity between them. We also note that the distribution of $\nu_{\tau}$ and $\bar{\nu}_{\tau}$ are also, in general, different from each other due to high-order corrections in neutrino-matter interactions (e.g., weak magnetism). This suggests that we need to analyze CFI with all six species of neutrinos distinguished. Otherwise some important features of CFI might be missed. We also note that stellar rotation can change the global matter distribution in the post-shock region. According to the study in \cite{Harada_2022}, FFC preferentially occurs in the equatorial region, since low values of $Y_e$ prevails there by the centrifugal force. This indicates that CFIs including the resonance-like mode may take place there. We deem the present work to be a necessary step to address these remaining issues and to quantify the impacts of CFI on CCSN dynamics.

\begin{acknowledgments}
This research used the K and Fugaku supercomputers provided by RIKEN, the FX10 provided by Tokyo University, the FX100 provided by Nagoya University, the Grand Chariot provided by Hokkaido University, and Oakforest-PACS provided by JCAHPC through the HPCI System Research Project (Project ID: hp130025, 140211, 150225, 150262, 160071, 160211, 170031, 170230, 170304, 180111, 180179, 180239, 190100, 190160, 200102, 200124, 210050, 210051, 210164, 220047, 220173, 220047, 220223 and 230033), and the Cray XC50 at Center for Computational Astrophysics, National Astronomical Observatory of Japan (NAOJ). This work is supported by 
Grant-in-Aid for Scientific Research
(19K03837, 20H01905,21H01083)
and 
Grant-in-Aid for Scientific Research on Innovative areas 
"Gravitational wave physics and astronomy:Genesis" 
(17H06357, 17H06365) and ”Unraveling the History of the Universe and Matter Evolution with Underground Physics” (19H05802 and 19H05811)
from the Ministry of Education, Culture, Sports, Science and Technology (MEXT), Japan.
For providing high performance computing resources, Computing Research Center, KEK, JLDG on SINET of NII, Research Center for Nuclear Physics, Osaka University, Yukawa Institute of Theoretical Physics, Kyoto University, Nagoya University, and Information Technology Center, University of Tokyo are acknowledged. This work was supported by 
MEXT as "Program for Promoting Researches on the Supercomputer Fugaku" 
(Toward a unified view of the universe: from large scale structures to planets, JPMXP1020200109) and the Particle, Nuclear and Astro Physics Simulation Program (Nos. 2020-004, 2021-004, 2022-003) of Institute of Particle and Nuclear Studies, High Energy Accelerator Research Organization (KEK).
RA is supported by JSPS Grant-in-Aid for JSPS Fellows (Grant No. 22J10298) from MEXT.
HN is supported by Grant-in Aid for Scientific Research (23K03468).
MZ is supported by JSPS Grant-in-Aid for JSPS Fellows (No. 22KJ2906) from MEXT.
SY is supported by Institute for Advanced Theoretical and Experimental Physics, Waseda University, and the Waseda University Grant for Special Research Projects (project No. 023C-141).
\end{acknowledgments}

\bibliography{references.bib} 

\providecommand{\noopsort}[1]{}\providecommand{\singleletter}[1]{#1}%
\begin{thebibliography}{87}%
\makeatletter
\providecommand \@ifxundefined [1]{%
 \@ifx{#1\undefined}
}%
\providecommand \@ifnum [1]{%
 \ifnum #1\expandafter \@firstoftwo
 \else \expandafter \@secondoftwo
 \fi
}%
\providecommand \@ifx [1]{%
 \ifx #1\expandafter \@firstoftwo
 \else \expandafter \@secondoftwo
 \fi
}%
\providecommand \natexlab [1]{#1}%
\providecommand \enquote  [1]{``#1''}%
\providecommand \bibnamefont  [1]{#1}%
\providecommand \bibfnamefont [1]{#1}%
\providecommand \citenamefont [1]{#1}%
\providecommand \href@noop [0]{\@secondoftwo}%
\providecommand \href [0]{\begingroup \@sanitize@url \@href}%
\providecommand \@href[1]{\@@startlink{#1}\@@href}%
\providecommand \@@href[1]{\endgroup#1\@@endlink}%
\providecommand \@sanitize@url [0]{\catcode `\\12\catcode `\$12\catcode
  `\&12\catcode `\#12\catcode `\^12\catcode `\_12\catcode `\%12\relax}%
\providecommand \@@startlink[1]{}%
\providecommand \@@endlink[0]{}%
\providecommand \url  [0]{\begingroup\@sanitize@url \@url }%
\providecommand \@url [1]{\endgroup\@href {#1}{\urlprefix }}%
\providecommand \urlprefix  [0]{URL }%
\providecommand \Eprint [0]{\href }%
\providecommand \doibase [0]{https://doi.org/}%
\providecommand \selectlanguage [0]{\@gobble}%
\providecommand \bibinfo  [0]{\@secondoftwo}%
\providecommand \bibfield  [0]{\@secondoftwo}%
\providecommand \translation [1]{[#1]}%
\providecommand \BibitemOpen [0]{}%
\providecommand \bibitemStop [0]{}%
\providecommand \bibitemNoStop [0]{.\EOS\space}%
\providecommand \EOS [0]{\spacefactor3000\relax}%
\providecommand \BibitemShut  [1]{\csname bibitem#1\endcsname}%
\let\auto@bib@innerbib\@empty
\bibitem [{\citenamefont {Ott}\ \emph {et~al.}(2018)\citenamefont {Ott},
  \citenamefont {Roberts}, \citenamefont {da~Silva~Schneider}, \citenamefont
  {Fedrow}, \citenamefont {Haas},\ and\ \citenamefont {Schnetter}}]{Ott_2018}%
  \BibitemOpen
  \bibfield  {author} {\bibinfo {author} {\bibfnamefont {C.~D.}\ \bibnamefont
  {Ott}}, \bibinfo {author} {\bibfnamefont {L.~F.}\ \bibnamefont {Roberts}},
  \bibinfo {author} {\bibfnamefont {A.}~\bibnamefont {da~Silva~Schneider}},
  \bibinfo {author} {\bibfnamefont {J.~M.}\ \bibnamefont {Fedrow}}, \bibinfo
  {author} {\bibfnamefont {R.}~\bibnamefont {Haas}},\ and\ \bibinfo {author}
  {\bibfnamefont {E.}~\bibnamefont {Schnetter}},\ }\href
  {https://doi.org/10.3847/2041-8213/aaa967} {\bibfield  {journal} {\bibinfo
  {journal} {The Astrophysical Journal Letters}\ }\textbf {\bibinfo {volume}
  {855}},\ \bibinfo {pages} {L3} (\bibinfo {year} {2018})}\BibitemShut
  {NoStop}%
\bibitem [{\citenamefont {Burrows}\ \emph {et~al.}(2019)\citenamefont
  {Burrows}, \citenamefont {Radice}, \citenamefont {Vartanyan}, \citenamefont
  {Nagakura}, \citenamefont {Skinner},\ and\ \citenamefont
  {Dolence}}]{10.1093/mnras/stz3223}%
  \BibitemOpen
  \bibfield  {author} {\bibinfo {author} {\bibfnamefont {A.}~\bibnamefont
  {Burrows}}, \bibinfo {author} {\bibfnamefont {D.}~\bibnamefont {Radice}},
  \bibinfo {author} {\bibfnamefont {D.}~\bibnamefont {Vartanyan}}, \bibinfo
  {author} {\bibfnamefont {H.}~\bibnamefont {Nagakura}}, \bibinfo {author}
  {\bibfnamefont {M.~A.}\ \bibnamefont {Skinner}},\ and\ \bibinfo {author}
  {\bibfnamefont {J.~C.}\ \bibnamefont {Dolence}},\ }\href
  {https://doi.org/10.1093/mnras/stz3223} {\bibfield  {journal} {\bibinfo
  {journal} {Monthly Notices of the Royal Astronomical Society}\ }\textbf
  {\bibinfo {volume} {491}},\ \bibinfo {pages} {2715} (\bibinfo {year}
  {2019})},\ \Eprint
  {https://arxiv.org/abs/https://academic.oup.com/mnras/article-pdf/491/2/2715/31221715/stz3223.pdf}
  {https://academic.oup.com/mnras/article-pdf/491/2/2715/31221715/stz3223.pdf}
  \BibitemShut {NoStop}%
\bibitem [{\citenamefont {Powell}\ and\ \citenamefont
  {Müller}(2020)}]{10.1093/mnras/staa1048}%
  \BibitemOpen
  \bibfield  {author} {\bibinfo {author} {\bibfnamefont {J.}~\bibnamefont
  {Powell}}\ and\ \bibinfo {author} {\bibfnamefont {B.}~\bibnamefont
  {Müller}},\ }\href {https://doi.org/10.1093/mnras/staa1048} {\bibfield
  {journal} {\bibinfo  {journal} {Monthly Notices of the Royal Astronomical
  Society}\ }\textbf {\bibinfo {volume} {494}},\ \bibinfo {pages} {4665}
  (\bibinfo {year} {2020})},\ \Eprint
  {https://arxiv.org/abs/https://academic.oup.com/mnras/article-pdf/494/4/4665/33174338/staa1048.pdf}
  {https://academic.oup.com/mnras/article-pdf/494/4/4665/33174338/staa1048.pdf}
  \BibitemShut {NoStop}%
\bibitem [{\citenamefont {Harada}\ \emph {et~al.}(2020)\citenamefont {Harada},
  \citenamefont {Nagakura}, \citenamefont {Iwakami}, \citenamefont {Okawa},
  \citenamefont {Furusawa}, \citenamefont {Sumiyoshi}, \citenamefont
  {Matsufuru},\ and\ \citenamefont {Yamada}}]{Harada_2020}%
  \BibitemOpen
  \bibfield  {author} {\bibinfo {author} {\bibfnamefont {A.}~\bibnamefont
  {Harada}}, \bibinfo {author} {\bibfnamefont {H.}~\bibnamefont {Nagakura}},
  \bibinfo {author} {\bibfnamefont {W.}~\bibnamefont {Iwakami}}, \bibinfo
  {author} {\bibfnamefont {H.}~\bibnamefont {Okawa}}, \bibinfo {author}
  {\bibfnamefont {S.}~\bibnamefont {Furusawa}}, \bibinfo {author}
  {\bibfnamefont {K.}~\bibnamefont {Sumiyoshi}}, \bibinfo {author}
  {\bibfnamefont {H.}~\bibnamefont {Matsufuru}},\ and\ \bibinfo {author}
  {\bibfnamefont {S.}~\bibnamefont {Yamada}},\ }\href
  {https://doi.org/10.3847/1538-4357/abb5a9} {\bibfield  {journal} {\bibinfo
  {journal} {The Astrophysical Journal}\ }\textbf {\bibinfo {volume} {902}},\
  \bibinfo {pages} {150} (\bibinfo {year} {2020})}\BibitemShut {NoStop}%
\bibitem [{\citenamefont {Bollig}\ \emph {et~al.}(2021)\citenamefont {Bollig},
  \citenamefont {Yadav}, \citenamefont {Kresse}, \citenamefont {Janka},
  \citenamefont {Müller},\ and\ \citenamefont {Heger}}]{Bollig_2021}%
  \BibitemOpen
  \bibfield  {author} {\bibinfo {author} {\bibfnamefont {R.}~\bibnamefont
  {Bollig}}, \bibinfo {author} {\bibfnamefont {N.}~\bibnamefont {Yadav}},
  \bibinfo {author} {\bibfnamefont {D.}~\bibnamefont {Kresse}}, \bibinfo
  {author} {\bibfnamefont {H.-T.}\ \bibnamefont {Janka}}, \bibinfo {author}
  {\bibfnamefont {B.}~\bibnamefont {Müller}},\ and\ \bibinfo {author}
  {\bibfnamefont {A.}~\bibnamefont {Heger}},\ }\href
  {https://doi.org/10.3847/1538-4357/abf82e} {\bibfield  {journal} {\bibinfo
  {journal} {The Astrophysical Journal}\ }\textbf {\bibinfo {volume} {915}},\
  \bibinfo {pages} {28} (\bibinfo {year} {2021})}\BibitemShut {NoStop}%
\bibitem [{\citenamefont {Nakamura}\ \emph {et~al.}(2022)\citenamefont
  {Nakamura}, \citenamefont {Takiwaki},\ and\ \citenamefont
  {Kotake}}]{10.1093/mnras/stac1586}%
  \BibitemOpen
  \bibfield  {author} {\bibinfo {author} {\bibfnamefont {K.}~\bibnamefont
  {Nakamura}}, \bibinfo {author} {\bibfnamefont {T.}~\bibnamefont {Takiwaki}},\
  and\ \bibinfo {author} {\bibfnamefont {K.}~\bibnamefont {Kotake}},\ }\href
  {https://doi.org/10.1093/mnras/stac1586} {\bibfield  {journal} {\bibinfo
  {journal} {Monthly Notices of the Royal Astronomical Society}\ }\textbf
  {\bibinfo {volume} {514}},\ \bibinfo {pages} {3941} (\bibinfo {year}
  {2022})},\ \Eprint
  {https://arxiv.org/abs/https://academic.oup.com/mnras/article-pdf/514/3/3941/44276121/stac1586.pdf}
  {https://academic.oup.com/mnras/article-pdf/514/3/3941/44276121/stac1586.pdf}
  \BibitemShut {NoStop}%
\bibitem [{\citenamefont {Mezzacappa}\ \emph {et~al.}(2023)\citenamefont
  {Mezzacappa}, \citenamefont {Marronetti}, \citenamefont {Landfield},
  \citenamefont {Lentz}, \citenamefont {Murphy}, \citenamefont {Raphael~Hix},
  \citenamefont {Harris}, \citenamefont {Bruenn}, \citenamefont {Blondin},
  \citenamefont {Bronson~Messer}, \citenamefont {Casanova},\ and\ \citenamefont
  {Kronzer}}]{PhysRevD.107.043008}%
  \BibitemOpen
  \bibfield  {author} {\bibinfo {author} {\bibfnamefont {A.}~\bibnamefont
  {Mezzacappa}}, \bibinfo {author} {\bibfnamefont {P.}~\bibnamefont
  {Marronetti}}, \bibinfo {author} {\bibfnamefont {R.~E.}\ \bibnamefont
  {Landfield}}, \bibinfo {author} {\bibfnamefont {E.~J.}\ \bibnamefont
  {Lentz}}, \bibinfo {author} {\bibfnamefont {R.~D.}\ \bibnamefont {Murphy}},
  \bibinfo {author} {\bibfnamefont {W.}~\bibnamefont {Raphael~Hix}}, \bibinfo
  {author} {\bibfnamefont {J.~A.}\ \bibnamefont {Harris}}, \bibinfo {author}
  {\bibfnamefont {S.~W.}\ \bibnamefont {Bruenn}}, \bibinfo {author}
  {\bibfnamefont {J.~M.}\ \bibnamefont {Blondin}}, \bibinfo {author}
  {\bibfnamefont {O.~E.}\ \bibnamefont {Bronson~Messer}}, \bibinfo {author}
  {\bibfnamefont {J.}~\bibnamefont {Casanova}},\ and\ \bibinfo {author}
  {\bibfnamefont {L.~L.}\ \bibnamefont {Kronzer}},\ }\href
  {https://doi.org/10.1103/PhysRevD.107.043008} {\bibfield  {journal} {\bibinfo
   {journal} {Phys. Rev. D}\ }\textbf {\bibinfo {volume} {107}},\ \bibinfo
  {pages} {043008} (\bibinfo {year} {2023})}\BibitemShut {NoStop}%
\bibitem [{\citenamefont {Pantaleone}(1992{\natexlab{a}})}]{PANTALEONE1992128}%
  \BibitemOpen
  \bibfield  {author} {\bibinfo {author} {\bibfnamefont {J.}~\bibnamefont
  {Pantaleone}},\ }\href
  {https://doi.org/https://doi.org/10.1016/0370-2693(92)91887-F} {\bibfield
  {journal} {\bibinfo  {journal} {Physics Letters B}\ }\textbf {\bibinfo
  {volume} {287}},\ \bibinfo {pages} {128} (\bibinfo {year}
  {1992}{\natexlab{a}})}\BibitemShut {NoStop}%
\bibitem [{\citenamefont {Pantaleone}(1992{\natexlab{b}})}]{PhysRevD.46.510}%
  \BibitemOpen
  \bibfield  {author} {\bibinfo {author} {\bibfnamefont {J.}~\bibnamefont
  {Pantaleone}},\ }\href {https://doi.org/10.1103/PhysRevD.46.510} {\bibfield
  {journal} {\bibinfo  {journal} {Phys. Rev. D}\ }\textbf {\bibinfo {volume}
  {46}},\ \bibinfo {pages} {510} (\bibinfo {year}
  {1992}{\natexlab{b}})}\BibitemShut {NoStop}%
\bibitem [{\citenamefont {Sigl}\ and\ \citenamefont
  {Raffelt}(1993)}]{Sigl1993GeneralKD}%
  \BibitemOpen
  \bibfield  {author} {\bibinfo {author} {\bibfnamefont {G.}~\bibnamefont
  {Sigl}}\ and\ \bibinfo {author} {\bibfnamefont {G.}~\bibnamefont {Raffelt}},\
  }\href@noop {} {\bibfield  {journal} {\bibinfo  {journal} {Nuclear Physics}\
  }\textbf {\bibinfo {volume} {406}},\ \bibinfo {pages} {423} (\bibinfo {year}
  {1993})}\BibitemShut {NoStop}%
\bibitem [{\citenamefont {Duan}\ \emph {et~al.}(2010)\citenamefont {Duan},
  \citenamefont {Fuller},\ and\ \citenamefont
  {Qian}}]{doi:10.1146/annurev.nucl.012809.104524}%
  \BibitemOpen
  \bibfield  {author} {\bibinfo {author} {\bibfnamefont {H.}~\bibnamefont
  {Duan}}, \bibinfo {author} {\bibfnamefont {G.~M.}\ \bibnamefont {Fuller}},\
  and\ \bibinfo {author} {\bibfnamefont {Y.-Z.}\ \bibnamefont {Qian}},\ }\href
  {https://doi.org/10.1146/annurev.nucl.012809.104524} {\bibfield  {journal}
  {\bibinfo  {journal} {Annual Review of Nuclear and Particle Science}\
  }\textbf {\bibinfo {volume} {60}},\ \bibinfo {pages} {569} (\bibinfo {year}
  {2010})},\ \Eprint
  {https://arxiv.org/abs/https://doi.org/10.1146/annurev.nucl.012809.104524}
  {https://doi.org/10.1146/annurev.nucl.012809.104524} \BibitemShut {NoStop}%
\bibitem [{\citenamefont {Chakraborty}\ \emph {et~al.}(2016)\citenamefont
  {Chakraborty}, \citenamefont {Hansen}, \citenamefont {Izaguirre},\ and\
  \citenamefont {Raffelt}}]{CHAKRABORTY2016366}%
  \BibitemOpen
  \bibfield  {author} {\bibinfo {author} {\bibfnamefont {S.}~\bibnamefont
  {Chakraborty}}, \bibinfo {author} {\bibfnamefont {R.}~\bibnamefont {Hansen}},
  \bibinfo {author} {\bibfnamefont {I.}~\bibnamefont {Izaguirre}},\ and\
  \bibinfo {author} {\bibfnamefont {G.}~\bibnamefont {Raffelt}},\ }\href
  {https://doi.org/https://doi.org/10.1016/j.nuclphysb.2016.02.012} {\bibfield
  {journal} {\bibinfo  {journal} {Nuclear Physics B}\ }\textbf {\bibinfo
  {volume} {908}},\ \bibinfo {pages} {366} (\bibinfo {year} {2016})},\ \bibinfo
  {note} {neutrino Oscillations: Celebrating the Nobel Prize in Physics
  2015}\BibitemShut {NoStop}%
\bibitem [{\citenamefont {Tamborra}\ and\ \citenamefont
  {Shalgar}(2021)}]{doi:10.1146/annurev-nucl-102920-050505}%
  \BibitemOpen
  \bibfield  {author} {\bibinfo {author} {\bibfnamefont {I.}~\bibnamefont
  {Tamborra}}\ and\ \bibinfo {author} {\bibfnamefont {S.}~\bibnamefont
  {Shalgar}},\ }\href {https://doi.org/10.1146/annurev-nucl-102920-050505}
  {\bibfield  {journal} {\bibinfo  {journal} {Annual Review of Nuclear and
  Particle Science}\ }\textbf {\bibinfo {volume} {71}},\ \bibinfo {pages} {165}
  (\bibinfo {year} {2021})},\ \Eprint
  {https://arxiv.org/abs/https://doi.org/10.1146/annurev-nucl-102920-050505}
  {https://doi.org/10.1146/annurev-nucl-102920-050505} \BibitemShut {NoStop}%
\bibitem [{\citenamefont {Richers}\ and\ \citenamefont
  {Sen}(2022)}]{https://doi.org/10.48550/arxiv.2207.03561}%
  \BibitemOpen
  \bibfield  {author} {\bibinfo {author} {\bibfnamefont {S.}~\bibnamefont
  {Richers}}\ and\ \bibinfo {author} {\bibfnamefont {M.}~\bibnamefont {Sen}},\
  }\href {https://doi.org/10.48550/ARXIV.2207.03561} {\bibinfo {title} {Fast
  flavor transformations}} (\bibinfo {year} {2022})\BibitemShut {NoStop}%
\bibitem [{\citenamefont {Fiorillo}\ and\ \citenamefont
  {Raffelt}(2023{\natexlab{a}})}]{PhysRevD.107.043024}%
  \BibitemOpen
  \bibfield  {author} {\bibinfo {author} {\bibfnamefont {D.~F.~G.}\
  \bibnamefont {Fiorillo}}\ and\ \bibinfo {author} {\bibfnamefont {G.~G.}\
  \bibnamefont {Raffelt}},\ }\href
  {https://doi.org/10.1103/PhysRevD.107.043024} {\bibfield  {journal} {\bibinfo
   {journal} {Phys. Rev. D}\ }\textbf {\bibinfo {volume} {107}},\ \bibinfo
  {pages} {043024} (\bibinfo {year} {2023}{\natexlab{a}})}\BibitemShut
  {NoStop}%
\bibitem [{\citenamefont {Fiorillo}\ and\ \citenamefont
  {Raffelt}(2023{\natexlab{b}})}]{PhysRevD.107.123024}%
  \BibitemOpen
  \bibfield  {author} {\bibinfo {author} {\bibfnamefont {D.~F.~G.}\
  \bibnamefont {Fiorillo}}\ and\ \bibinfo {author} {\bibfnamefont {G.~G.}\
  \bibnamefont {Raffelt}},\ }\href
  {https://doi.org/10.1103/PhysRevD.107.123024} {\bibfield  {journal} {\bibinfo
   {journal} {Phys. Rev. D}\ }\textbf {\bibinfo {volume} {107}},\ \bibinfo
  {pages} {123024} (\bibinfo {year} {2023}{\natexlab{b}})}\BibitemShut
  {NoStop}%
\bibitem [{\citenamefont {Morinaga}(2022)}]{PhysRevD.105.L101301}%
  \BibitemOpen
  \bibfield  {author} {\bibinfo {author} {\bibfnamefont {T.}~\bibnamefont
  {Morinaga}},\ }\href {https://doi.org/10.1103/PhysRevD.105.L101301}
  {\bibfield  {journal} {\bibinfo  {journal} {Phys. Rev. D}\ }\textbf {\bibinfo
  {volume} {105}},\ \bibinfo {pages} {L101301} (\bibinfo {year}
  {2022})}\BibitemShut {NoStop}%
\bibitem [{\citenamefont {Tamborra}\ \emph {et~al.}(2017)\citenamefont
  {Tamborra}, \citenamefont {Hüdepohl}, \citenamefont {Raffelt},\ and\
  \citenamefont {Janka}}]{Tamborra_2017}%
  \BibitemOpen
  \bibfield  {author} {\bibinfo {author} {\bibfnamefont {I.}~\bibnamefont
  {Tamborra}}, \bibinfo {author} {\bibfnamefont {L.}~\bibnamefont {Hüdepohl}},
  \bibinfo {author} {\bibfnamefont {G.~G.}\ \bibnamefont {Raffelt}},\ and\
  \bibinfo {author} {\bibfnamefont {H.-T.}\ \bibnamefont {Janka}},\ }\href
  {https://doi.org/10.3847/1538-4357/aa6a18} {\bibfield  {journal} {\bibinfo
  {journal} {The Astrophysical Journal}\ }\textbf {\bibinfo {volume} {839}},\
  \bibinfo {pages} {132} (\bibinfo {year} {2017})}\BibitemShut {NoStop}%
\bibitem [{\citenamefont {Abbar}\ \emph {et~al.}(2019)\citenamefont {Abbar},
  \citenamefont {Duan}, \citenamefont {Sumiyoshi}, \citenamefont {Takiwaki},\
  and\ \citenamefont {Volpe}}]{PhysRevD.100.043004}%
  \BibitemOpen
  \bibfield  {author} {\bibinfo {author} {\bibfnamefont {S.}~\bibnamefont
  {Abbar}}, \bibinfo {author} {\bibfnamefont {H.}~\bibnamefont {Duan}},
  \bibinfo {author} {\bibfnamefont {K.}~\bibnamefont {Sumiyoshi}}, \bibinfo
  {author} {\bibfnamefont {T.}~\bibnamefont {Takiwaki}},\ and\ \bibinfo
  {author} {\bibfnamefont {M.~C.}\ \bibnamefont {Volpe}},\ }\href
  {https://doi.org/10.1103/PhysRevD.100.043004} {\bibfield  {journal} {\bibinfo
   {journal} {Phys. Rev. D}\ }\textbf {\bibinfo {volume} {100}},\ \bibinfo
  {pages} {043004} (\bibinfo {year} {2019})}\BibitemShut {NoStop}%
\bibitem [{\citenamefont {Nagakura}\ \emph
  {et~al.}(2019{\natexlab{a}})\citenamefont {Nagakura}, \citenamefont
  {Morinaga}, \citenamefont {Kato},\ and\ \citenamefont
  {Yamada}}]{Nagakura_2019}%
  \BibitemOpen
  \bibfield  {author} {\bibinfo {author} {\bibfnamefont {H.}~\bibnamefont
  {Nagakura}}, \bibinfo {author} {\bibfnamefont {T.}~\bibnamefont {Morinaga}},
  \bibinfo {author} {\bibfnamefont {C.}~\bibnamefont {Kato}},\ and\ \bibinfo
  {author} {\bibfnamefont {S.}~\bibnamefont {Yamada}},\ }\href
  {https://doi.org/10.3847/1538-4357/ab4cf2} {\bibfield  {journal} {\bibinfo
  {journal} {The Astrophysical Journal}\ }\textbf {\bibinfo {volume} {886}},\
  \bibinfo {pages} {139} (\bibinfo {year} {2019}{\natexlab{a}})}\BibitemShut
  {NoStop}%
\bibitem [{\citenamefont {Delfan~Azari}\ \emph {et~al.}(2020)\citenamefont
  {Delfan~Azari}, \citenamefont {Yamada}, \citenamefont {Morinaga},
  \citenamefont {Nagakura}, \citenamefont {Furusawa}, \citenamefont {Harada},
  \citenamefont {Okawa}, \citenamefont {Iwakami},\ and\ \citenamefont
  {Sumiyoshi}}]{PhysRevD.101.023018}%
  \BibitemOpen
  \bibfield  {author} {\bibinfo {author} {\bibfnamefont {M.}~\bibnamefont
  {Delfan~Azari}}, \bibinfo {author} {\bibfnamefont {S.}~\bibnamefont
  {Yamada}}, \bibinfo {author} {\bibfnamefont {T.}~\bibnamefont {Morinaga}},
  \bibinfo {author} {\bibfnamefont {H.}~\bibnamefont {Nagakura}}, \bibinfo
  {author} {\bibfnamefont {S.}~\bibnamefont {Furusawa}}, \bibinfo {author}
  {\bibfnamefont {A.}~\bibnamefont {Harada}}, \bibinfo {author} {\bibfnamefont
  {H.}~\bibnamefont {Okawa}}, \bibinfo {author} {\bibfnamefont
  {W.}~\bibnamefont {Iwakami}},\ and\ \bibinfo {author} {\bibfnamefont
  {K.}~\bibnamefont {Sumiyoshi}},\ }\href
  {https://doi.org/10.1103/PhysRevD.101.023018} {\bibfield  {journal} {\bibinfo
   {journal} {Phys. Rev. D}\ }\textbf {\bibinfo {volume} {101}},\ \bibinfo
  {pages} {023018} (\bibinfo {year} {2020})}\BibitemShut {NoStop}%
\bibitem [{\citenamefont {Abbar}\ \emph {et~al.}(2020)\citenamefont {Abbar},
  \citenamefont {Duan}, \citenamefont {Sumiyoshi}, \citenamefont {Takiwaki},\
  and\ \citenamefont {Volpe}}]{PhysRevD.101.043016}%
  \BibitemOpen
  \bibfield  {author} {\bibinfo {author} {\bibfnamefont {S.}~\bibnamefont
  {Abbar}}, \bibinfo {author} {\bibfnamefont {H.}~\bibnamefont {Duan}},
  \bibinfo {author} {\bibfnamefont {K.}~\bibnamefont {Sumiyoshi}}, \bibinfo
  {author} {\bibfnamefont {T.}~\bibnamefont {Takiwaki}},\ and\ \bibinfo
  {author} {\bibfnamefont {M.~C.}\ \bibnamefont {Volpe}},\ }\href
  {https://doi.org/10.1103/PhysRevD.101.043016} {\bibfield  {journal} {\bibinfo
   {journal} {Phys. Rev. D}\ }\textbf {\bibinfo {volume} {101}},\ \bibinfo
  {pages} {043016} (\bibinfo {year} {2020})}\BibitemShut {NoStop}%
\bibitem [{\citenamefont {Glas}\ \emph {et~al.}(2020)\citenamefont {Glas},
  \citenamefont {Janka}, \citenamefont {Capozzi}, \citenamefont {Sen},
  \citenamefont {Dasgupta}, \citenamefont {Mirizzi},\ and\ \citenamefont
  {Sigl}}]{PhysRevD.101.063001}%
  \BibitemOpen
  \bibfield  {author} {\bibinfo {author} {\bibfnamefont {R.}~\bibnamefont
  {Glas}}, \bibinfo {author} {\bibfnamefont {H.-T.}\ \bibnamefont {Janka}},
  \bibinfo {author} {\bibfnamefont {F.}~\bibnamefont {Capozzi}}, \bibinfo
  {author} {\bibfnamefont {M.}~\bibnamefont {Sen}}, \bibinfo {author}
  {\bibfnamefont {B.}~\bibnamefont {Dasgupta}}, \bibinfo {author}
  {\bibfnamefont {A.}~\bibnamefont {Mirizzi}},\ and\ \bibinfo {author}
  {\bibfnamefont {G.}~\bibnamefont {Sigl}},\ }\href
  {https://doi.org/10.1103/PhysRevD.101.063001} {\bibfield  {journal} {\bibinfo
   {journal} {Phys. Rev. D}\ }\textbf {\bibinfo {volume} {101}},\ \bibinfo
  {pages} {063001} (\bibinfo {year} {2020})}\BibitemShut {NoStop}%
\bibitem [{\citenamefont {Abbar}(2020)}]{Abbar_2020}%
  \BibitemOpen
  \bibfield  {author} {\bibinfo {author} {\bibfnamefont {S.}~\bibnamefont
  {Abbar}},\ }\href {https://doi.org/10.1088/1475-7516/2020/05/027} {\bibfield
  {journal} {\bibinfo  {journal} {Journal of Cosmology and Astroparticle
  Physics}\ }\textbf {\bibinfo {volume} {2020}}\bibinfo  {number} { (05)},\
  \bibinfo {pages} {027}}\BibitemShut {NoStop}%
\bibitem [{\citenamefont {Capozzi}\ \emph {et~al.}(2021)\citenamefont
  {Capozzi}, \citenamefont {Abbar}, \citenamefont {Bollig},\ and\ \citenamefont
  {Janka}}]{PhysRevD.103.063013}%
  \BibitemOpen
\bibfield  {number} {  }\bibfield  {author} {\bibinfo {author} {\bibfnamefont
  {F.}~\bibnamefont {Capozzi}}, \bibinfo {author} {\bibfnamefont
  {S.}~\bibnamefont {Abbar}}, \bibinfo {author} {\bibfnamefont
  {R.}~\bibnamefont {Bollig}},\ and\ \bibinfo {author} {\bibfnamefont {H.-T.}\
  \bibnamefont {Janka}},\ }\href {https://doi.org/10.1103/PhysRevD.103.063013}
  {\bibfield  {journal} {\bibinfo  {journal} {Phys. Rev. D}\ }\textbf {\bibinfo
  {volume} {103}},\ \bibinfo {pages} {063013} (\bibinfo {year}
  {2021})}\BibitemShut {NoStop}%
\bibitem [{\citenamefont {Nagakura}\ \emph {et~al.}(2021)\citenamefont
  {Nagakura}, \citenamefont {Burrows}, \citenamefont {Johns},\ and\
  \citenamefont {Fuller}}]{PhysRevD.104.083025}%
  \BibitemOpen
  \bibfield  {author} {\bibinfo {author} {\bibfnamefont {H.}~\bibnamefont
  {Nagakura}}, \bibinfo {author} {\bibfnamefont {A.}~\bibnamefont {Burrows}},
  \bibinfo {author} {\bibfnamefont {L.}~\bibnamefont {Johns}},\ and\ \bibinfo
  {author} {\bibfnamefont {G.~M.}\ \bibnamefont {Fuller}},\ }\href
  {https://doi.org/10.1103/PhysRevD.104.083025} {\bibfield  {journal} {\bibinfo
   {journal} {Phys. Rev. D}\ }\textbf {\bibinfo {volume} {104}},\ \bibinfo
  {pages} {083025} (\bibinfo {year} {2021})}\BibitemShut {NoStop}%
\bibitem [{\citenamefont {Harada}\ and\ \citenamefont
  {Nagakura}(2022)}]{Harada_2022}%
  \BibitemOpen
  \bibfield  {author} {\bibinfo {author} {\bibfnamefont {A.}~\bibnamefont
  {Harada}}\ and\ \bibinfo {author} {\bibfnamefont {H.}~\bibnamefont
  {Nagakura}},\ }\href {https://doi.org/10.3847/1538-4357/ac38a0} {\bibfield
  {journal} {\bibinfo  {journal} {The Astrophysical Journal}\ }\textbf
  {\bibinfo {volume} {924}},\ \bibinfo {pages} {109} (\bibinfo {year}
  {2022})}\BibitemShut {NoStop}%
\bibitem [{\citenamefont {Akaho}\ \emph {et~al.}(2023)\citenamefont {Akaho},
  \citenamefont {Harada}, \citenamefont {Nagakura}, \citenamefont {Iwakami},
  \citenamefont {Okawa}, \citenamefont {Furusawa}, \citenamefont {Matsufuru},
  \citenamefont {Sumiyoshi},\ and\ \citenamefont {Yamada}}]{Akaho_2023}%
  \BibitemOpen
  \bibfield  {author} {\bibinfo {author} {\bibfnamefont {R.}~\bibnamefont
  {Akaho}}, \bibinfo {author} {\bibfnamefont {A.}~\bibnamefont {Harada}},
  \bibinfo {author} {\bibfnamefont {H.}~\bibnamefont {Nagakura}}, \bibinfo
  {author} {\bibfnamefont {W.}~\bibnamefont {Iwakami}}, \bibinfo {author}
  {\bibfnamefont {H.}~\bibnamefont {Okawa}}, \bibinfo {author} {\bibfnamefont
  {S.}~\bibnamefont {Furusawa}}, \bibinfo {author} {\bibfnamefont
  {H.}~\bibnamefont {Matsufuru}}, \bibinfo {author} {\bibfnamefont
  {K.}~\bibnamefont {Sumiyoshi}},\ and\ \bibinfo {author} {\bibfnamefont
  {S.}~\bibnamefont {Yamada}},\ }\href
  {https://doi.org/10.3847/1538-4357/acad76} {\bibfield  {journal} {\bibinfo
  {journal} {The Astrophysical Journal}\ }\textbf {\bibinfo {volume} {944}},\
  \bibinfo {pages} {60} (\bibinfo {year} {2023})}\BibitemShut {NoStop}%
\bibitem [{\citenamefont {Delfan~Azari}\ \emph {et~al.}(2019)\citenamefont
  {Delfan~Azari}, \citenamefont {Yamada}, \citenamefont {Morinaga},
  \citenamefont {Iwakami}, \citenamefont {Okawa}, \citenamefont {Nagakura},\
  and\ \citenamefont {Sumiyoshi}}]{PhysRevD.99.103011}%
  \BibitemOpen
  \bibfield  {author} {\bibinfo {author} {\bibfnamefont {M.}~\bibnamefont
  {Delfan~Azari}}, \bibinfo {author} {\bibfnamefont {S.}~\bibnamefont
  {Yamada}}, \bibinfo {author} {\bibfnamefont {T.}~\bibnamefont {Morinaga}},
  \bibinfo {author} {\bibfnamefont {W.}~\bibnamefont {Iwakami}}, \bibinfo
  {author} {\bibfnamefont {H.}~\bibnamefont {Okawa}}, \bibinfo {author}
  {\bibfnamefont {H.}~\bibnamefont {Nagakura}},\ and\ \bibinfo {author}
  {\bibfnamefont {K.}~\bibnamefont {Sumiyoshi}},\ }\href
  {https://doi.org/10.1103/PhysRevD.99.103011} {\bibfield  {journal} {\bibinfo
  {journal} {Phys. Rev. D}\ }\textbf {\bibinfo {volume} {99}},\ \bibinfo
  {pages} {103011} (\bibinfo {year} {2019})}\BibitemShut {NoStop}%
\bibitem [{\citenamefont {Johns}\ \emph
  {et~al.}(2020{\natexlab{a}})\citenamefont {Johns}, \citenamefont {Nagakura},
  \citenamefont {Fuller},\ and\ \citenamefont {Burrows}}]{PhysRevD.101.043009}%
  \BibitemOpen
  \bibfield  {author} {\bibinfo {author} {\bibfnamefont {L.}~\bibnamefont
  {Johns}}, \bibinfo {author} {\bibfnamefont {H.}~\bibnamefont {Nagakura}},
  \bibinfo {author} {\bibfnamefont {G.~M.}\ \bibnamefont {Fuller}},\ and\
  \bibinfo {author} {\bibfnamefont {A.}~\bibnamefont {Burrows}},\ }\href
  {https://doi.org/10.1103/PhysRevD.101.043009} {\bibfield  {journal} {\bibinfo
   {journal} {Phys. Rev. D}\ }\textbf {\bibinfo {volume} {101}},\ \bibinfo
  {pages} {043009} (\bibinfo {year} {2020}{\natexlab{a}})}\BibitemShut
  {NoStop}%
\bibitem [{\citenamefont {Johns}\ \emph
  {et~al.}(2020{\natexlab{b}})\citenamefont {Johns}, \citenamefont {Nagakura},
  \citenamefont {Fuller},\ and\ \citenamefont {Burrows}}]{PhysRevD.102.103017}%
  \BibitemOpen
  \bibfield  {author} {\bibinfo {author} {\bibfnamefont {L.}~\bibnamefont
  {Johns}}, \bibinfo {author} {\bibfnamefont {H.}~\bibnamefont {Nagakura}},
  \bibinfo {author} {\bibfnamefont {G.~M.}\ \bibnamefont {Fuller}},\ and\
  \bibinfo {author} {\bibfnamefont {A.}~\bibnamefont {Burrows}},\ }\href
  {https://doi.org/10.1103/PhysRevD.102.103017} {\bibfield  {journal} {\bibinfo
   {journal} {Phys. Rev. D}\ }\textbf {\bibinfo {volume} {102}},\ \bibinfo
  {pages} {103017} (\bibinfo {year} {2020}{\natexlab{b}})}\BibitemShut
  {NoStop}%
\bibitem [{\citenamefont {Wu}\ \emph {et~al.}(2021)\citenamefont {Wu},
  \citenamefont {George}, \citenamefont {Lin},\ and\ \citenamefont
  {Xiong}}]{PhysRevD.104.103003}%
  \BibitemOpen
  \bibfield  {author} {\bibinfo {author} {\bibfnamefont {M.-R.}\ \bibnamefont
  {Wu}}, \bibinfo {author} {\bibfnamefont {M.}~\bibnamefont {George}}, \bibinfo
  {author} {\bibfnamefont {C.-Y.}\ \bibnamefont {Lin}},\ and\ \bibinfo {author}
  {\bibfnamefont {Z.}~\bibnamefont {Xiong}},\ }\href
  {https://doi.org/10.1103/PhysRevD.104.103003} {\bibfield  {journal} {\bibinfo
   {journal} {Phys. Rev. D}\ }\textbf {\bibinfo {volume} {104}},\ \bibinfo
  {pages} {103003} (\bibinfo {year} {2021})}\BibitemShut {NoStop}%
\bibitem [{\citenamefont {Padilla-Gay}\ \emph
  {et~al.}(2022{\natexlab{a}})\citenamefont {Padilla-Gay}, \citenamefont
  {Tamborra},\ and\ \citenamefont {Raffelt}}]{PhysRevLett.128.121102}%
  \BibitemOpen
  \bibfield  {author} {\bibinfo {author} {\bibfnamefont {I.}~\bibnamefont
  {Padilla-Gay}}, \bibinfo {author} {\bibfnamefont {I.}~\bibnamefont
  {Tamborra}},\ and\ \bibinfo {author} {\bibfnamefont {G.~G.}\ \bibnamefont
  {Raffelt}},\ }\href {https://doi.org/10.1103/PhysRevLett.128.121102}
  {\bibfield  {journal} {\bibinfo  {journal} {Phys. Rev. Lett.}\ }\textbf
  {\bibinfo {volume} {128}},\ \bibinfo {pages} {121102} (\bibinfo {year}
  {2022}{\natexlab{a}})}\BibitemShut {NoStop}%
\bibitem [{\citenamefont {Richers}\ \emph {et~al.}(2021)\citenamefont
  {Richers}, \citenamefont {Willcox},\ and\ \citenamefont
  {Ford}}]{PhysRevD.104.103023}%
  \BibitemOpen
  \bibfield  {author} {\bibinfo {author} {\bibfnamefont {S.}~\bibnamefont
  {Richers}}, \bibinfo {author} {\bibfnamefont {D.}~\bibnamefont {Willcox}},\
  and\ \bibinfo {author} {\bibfnamefont {N.}~\bibnamefont {Ford}},\ }\href
  {https://doi.org/10.1103/PhysRevD.104.103023} {\bibfield  {journal} {\bibinfo
   {journal} {Phys. Rev. D}\ }\textbf {\bibinfo {volume} {104}},\ \bibinfo
  {pages} {103023} (\bibinfo {year} {2021})}\BibitemShut {NoStop}%
\bibitem [{\citenamefont {Bhattacharyya}\ and\ \citenamefont
  {Dasgupta}(2022)}]{PhysRevD.106.103039}%
  \BibitemOpen
  \bibfield  {author} {\bibinfo {author} {\bibfnamefont {S.}~\bibnamefont
  {Bhattacharyya}}\ and\ \bibinfo {author} {\bibfnamefont {B.}~\bibnamefont
  {Dasgupta}},\ }\href {https://doi.org/10.1103/PhysRevD.106.103039} {\bibfield
   {journal} {\bibinfo  {journal} {Phys. Rev. D}\ }\textbf {\bibinfo {volume}
  {106}},\ \bibinfo {pages} {103039} (\bibinfo {year} {2022})}\BibitemShut
  {NoStop}%
\bibitem [{\citenamefont {Nagakura}\ and\ \citenamefont
  {Zaizen}(2022)}]{PhysRevLett.129.261101}%
  \BibitemOpen
  \bibfield  {author} {\bibinfo {author} {\bibfnamefont {H.}~\bibnamefont
  {Nagakura}}\ and\ \bibinfo {author} {\bibfnamefont {M.}~\bibnamefont
  {Zaizen}},\ }\href {https://doi.org/10.1103/PhysRevLett.129.261101}
  {\bibfield  {journal} {\bibinfo  {journal} {Phys. Rev. Lett.}\ }\textbf
  {\bibinfo {volume} {129}},\ \bibinfo {pages} {261101} (\bibinfo {year}
  {2022})}\BibitemShut {NoStop}%
\bibitem [{\citenamefont {Zaizen}\ and\ \citenamefont
  {Nagakura}(2023{\natexlab{a}})}]{PhysRevD.107.103022}%
  \BibitemOpen
  \bibfield  {author} {\bibinfo {author} {\bibfnamefont {M.}~\bibnamefont
  {Zaizen}}\ and\ \bibinfo {author} {\bibfnamefont {H.}~\bibnamefont
  {Nagakura}},\ }\href {https://doi.org/10.1103/PhysRevD.107.103022} {\bibfield
   {journal} {\bibinfo  {journal} {Phys. Rev. D}\ }\textbf {\bibinfo {volume}
  {107}},\ \bibinfo {pages} {103022} (\bibinfo {year}
  {2023}{\natexlab{a}})}\BibitemShut {NoStop}%
\bibitem [{\citenamefont {Zaizen}\ and\ \citenamefont
  {Nagakura}(2023{\natexlab{b}})}]{PhysRevD.107.123021}%
  \BibitemOpen
  \bibfield  {author} {\bibinfo {author} {\bibfnamefont {M.}~\bibnamefont
  {Zaizen}}\ and\ \bibinfo {author} {\bibfnamefont {H.}~\bibnamefont
  {Nagakura}},\ }\href {https://doi.org/10.1103/PhysRevD.107.123021} {\bibfield
   {journal} {\bibinfo  {journal} {Phys. Rev. D}\ }\textbf {\bibinfo {volume}
  {107}},\ \bibinfo {pages} {123021} (\bibinfo {year}
  {2023}{\natexlab{b}})}\BibitemShut {NoStop}%
\bibitem [{\citenamefont {Xiong}\ \emph
  {et~al.}(2023{\natexlab{a}})\citenamefont {Xiong}, \citenamefont {Wu},
  \citenamefont {Abbar}, \citenamefont {Bhattacharyya}, \citenamefont
  {George},\ and\ \citenamefont {Lin}}]{xiong2023evaluating}%
  \BibitemOpen
  \bibfield  {author} {\bibinfo {author} {\bibfnamefont {Z.}~\bibnamefont
  {Xiong}}, \bibinfo {author} {\bibfnamefont {M.-R.}\ \bibnamefont {Wu}},
  \bibinfo {author} {\bibfnamefont {S.}~\bibnamefont {Abbar}}, \bibinfo
  {author} {\bibfnamefont {S.}~\bibnamefont {Bhattacharyya}}, \bibinfo {author}
  {\bibfnamefont {M.}~\bibnamefont {George}},\ and\ \bibinfo {author}
  {\bibfnamefont {C.-Y.}\ \bibnamefont {Lin}},\ }\href@noop {} {\bibinfo
  {title} {Evaluating approximate asymptotic distributions for fast neutrino
  flavor conversions in a periodic 1d box}} (\bibinfo {year}
  {2023}{\natexlab{a}}),\ \Eprint {https://arxiv.org/abs/2307.11129}
  {arXiv:2307.11129 [astro-ph.HE]} \BibitemShut {NoStop}%
\bibitem [{\citenamefont {Martin}\ \emph {et~al.}(2021)\citenamefont {Martin},
  \citenamefont {Carlson}, \citenamefont {Cirigliano},\ and\ \citenamefont
  {Duan}}]{PhysRevD.103.063001}%
  \BibitemOpen
  \bibfield  {author} {\bibinfo {author} {\bibfnamefont {J.~D.}\ \bibnamefont
  {Martin}}, \bibinfo {author} {\bibfnamefont {J.}~\bibnamefont {Carlson}},
  \bibinfo {author} {\bibfnamefont {V.}~\bibnamefont {Cirigliano}},\ and\
  \bibinfo {author} {\bibfnamefont {H.}~\bibnamefont {Duan}},\ }\href
  {https://doi.org/10.1103/PhysRevD.103.063001} {\bibfield  {journal} {\bibinfo
   {journal} {Phys. Rev. D}\ }\textbf {\bibinfo {volume} {103}},\ \bibinfo
  {pages} {063001} (\bibinfo {year} {2021})}\BibitemShut {NoStop}%
\bibitem [{\citenamefont {Sigl}(2022)}]{PhysRevD.105.043005}%
  \BibitemOpen
  \bibfield  {author} {\bibinfo {author} {\bibfnamefont {G.}~\bibnamefont
  {Sigl}},\ }\href {https://doi.org/10.1103/PhysRevD.105.043005} {\bibfield
  {journal} {\bibinfo  {journal} {Phys. Rev. D}\ }\textbf {\bibinfo {volume}
  {105}},\ \bibinfo {pages} {043005} (\bibinfo {year} {2022})}\BibitemShut
  {NoStop}%
\bibitem [{\citenamefont {Johns}\ and\ \citenamefont
  {Nagakura}(2022)}]{PhysRevD.106.043031}%
  \BibitemOpen
  \bibfield  {author} {\bibinfo {author} {\bibfnamefont {L.}~\bibnamefont
  {Johns}}\ and\ \bibinfo {author} {\bibfnamefont {H.}~\bibnamefont
  {Nagakura}},\ }\href {https://doi.org/10.1103/PhysRevD.106.043031} {\bibfield
   {journal} {\bibinfo  {journal} {Phys. Rev. D}\ }\textbf {\bibinfo {volume}
  {106}},\ \bibinfo {pages} {043031} (\bibinfo {year} {2022})}\BibitemShut
  {NoStop}%
\bibitem [{\citenamefont {Shalgar}\ and\ \citenamefont
  {Tamborra}(2021)}]{PhysRevD.103.063002}%
  \BibitemOpen
  \bibfield  {author} {\bibinfo {author} {\bibfnamefont {S.}~\bibnamefont
  {Shalgar}}\ and\ \bibinfo {author} {\bibfnamefont {I.}~\bibnamefont
  {Tamborra}},\ }\href {https://doi.org/10.1103/PhysRevD.103.063002} {\bibfield
   {journal} {\bibinfo  {journal} {Phys. Rev. D}\ }\textbf {\bibinfo {volume}
  {103}},\ \bibinfo {pages} {063002} (\bibinfo {year} {2021})}\BibitemShut
  {NoStop}%
\bibitem [{\citenamefont {Kato}\ \emph {et~al.}(2021)\citenamefont {Kato},
  \citenamefont {Nagakura},\ and\ \citenamefont {Morinaga}}]{Kato_2021}%
  \BibitemOpen
  \bibfield  {author} {\bibinfo {author} {\bibfnamefont {C.}~\bibnamefont
  {Kato}}, \bibinfo {author} {\bibfnamefont {H.}~\bibnamefont {Nagakura}},\
  and\ \bibinfo {author} {\bibfnamefont {T.}~\bibnamefont {Morinaga}},\ }\href
  {https://doi.org/10.3847/1538-4365/ac2aa4} {\bibfield  {journal} {\bibinfo
  {journal} {The Astrophysical Journal Supplement Series}\ }\textbf {\bibinfo
  {volume} {257}},\ \bibinfo {pages} {55} (\bibinfo {year} {2021})}\BibitemShut
  {NoStop}%
\bibitem [{\citenamefont {Sasaki}\ and\ \citenamefont
  {Takiwaki}(2022)}]{10.1093/ptep/ptac082}%
  \BibitemOpen
  \bibfield  {author} {\bibinfo {author} {\bibfnamefont {H.}~\bibnamefont
  {Sasaki}}\ and\ \bibinfo {author} {\bibfnamefont {T.}~\bibnamefont
  {Takiwaki}},\ }\bibfield  {journal} {\bibinfo  {journal} {Progress of
  Theoretical and Experimental Physics}\ }\textbf {\bibinfo {volume} {2022}},\
  \href {https://doi.org/10.1093/ptep/ptac082} {10.1093/ptep/ptac082} (\bibinfo
  {year} {2022}),\ \bibinfo {note} {073E01},\ \Eprint
  {https://arxiv.org/abs/https://academic.oup.com/ptep/article-pdf/2022/7/073E01/44400755/ptac082.pdf}
  {https://academic.oup.com/ptep/article-pdf/2022/7/073E01/44400755/ptac082.pdf}
  \BibitemShut {NoStop}%
\bibitem [{\citenamefont {Hansen}\ \emph {et~al.}(2022)\citenamefont {Hansen},
  \citenamefont {Shalgar},\ and\ \citenamefont
  {Tamborra}}]{PhysRevD.105.123003}%
  \BibitemOpen
  \bibfield  {author} {\bibinfo {author} {\bibfnamefont {R.~S.~L.}\
  \bibnamefont {Hansen}}, \bibinfo {author} {\bibfnamefont {S.}~\bibnamefont
  {Shalgar}},\ and\ \bibinfo {author} {\bibfnamefont {I.}~\bibnamefont
  {Tamborra}},\ }\href {https://doi.org/10.1103/PhysRevD.105.123003} {\bibfield
   {journal} {\bibinfo  {journal} {Phys. Rev. D}\ }\textbf {\bibinfo {volume}
  {105}},\ \bibinfo {pages} {123003} (\bibinfo {year} {2022})}\BibitemShut
  {NoStop}%
\bibitem [{\citenamefont {Kato}\ and\ \citenamefont
  {Nagakura}(2022)}]{Kato_2022}%
  \BibitemOpen
  \bibfield  {author} {\bibinfo {author} {\bibfnamefont {C.}~\bibnamefont
  {Kato}}\ and\ \bibinfo {author} {\bibfnamefont {H.}~\bibnamefont
  {Nagakura}},\ }\bibfield  {journal} {\bibinfo  {journal} {Physical Review D}\
  }\textbf {\bibinfo {volume} {106}},\ \href
  {https://doi.org/10.1103/physrevd.106.123013} {10.1103/physrevd.106.123013}
  (\bibinfo {year} {2022})\BibitemShut {NoStop}%
\bibitem [{\citenamefont {Padilla-Gay}\ \emph
  {et~al.}(2022{\natexlab{b}})\citenamefont {Padilla-Gay}, \citenamefont
  {Tamborra},\ and\ \citenamefont {Raffelt}}]{PhysRevD.106.103031}%
  \BibitemOpen
  \bibfield  {author} {\bibinfo {author} {\bibfnamefont {I.}~\bibnamefont
  {Padilla-Gay}}, \bibinfo {author} {\bibfnamefont {I.}~\bibnamefont
  {Tamborra}},\ and\ \bibinfo {author} {\bibfnamefont {G.~G.}\ \bibnamefont
  {Raffelt}},\ }\href {https://doi.org/10.1103/PhysRevD.106.103031} {\bibfield
  {journal} {\bibinfo  {journal} {Phys. Rev. D}\ }\textbf {\bibinfo {volume}
  {106}},\ \bibinfo {pages} {103031} (\bibinfo {year}
  {2022}{\natexlab{b}})}\BibitemShut {NoStop}%
\bibitem [{\citenamefont {Capozzi}\ \emph {et~al.}(2019)\citenamefont
  {Capozzi}, \citenamefont {Dasgupta}, \citenamefont {Mirizzi}, \citenamefont
  {Sen},\ and\ \citenamefont {Sigl}}]{PhysRevLett.122.091101}%
  \BibitemOpen
  \bibfield  {author} {\bibinfo {author} {\bibfnamefont {F.}~\bibnamefont
  {Capozzi}}, \bibinfo {author} {\bibfnamefont {B.}~\bibnamefont {Dasgupta}},
  \bibinfo {author} {\bibfnamefont {A.}~\bibnamefont {Mirizzi}}, \bibinfo
  {author} {\bibfnamefont {M.}~\bibnamefont {Sen}},\ and\ \bibinfo {author}
  {\bibfnamefont {G.}~\bibnamefont {Sigl}},\ }\href
  {https://doi.org/10.1103/PhysRevLett.122.091101} {\bibfield  {journal}
  {\bibinfo  {journal} {Phys. Rev. Lett.}\ }\textbf {\bibinfo {volume} {122}},\
  \bibinfo {pages} {091101} (\bibinfo {year} {2019})}\BibitemShut {NoStop}%
\bibitem [{\citenamefont {Shalgar}\ and\ \citenamefont
  {Tamborra}(2023{\natexlab{a}})}]{PhysRevD.107.063025}%
  \BibitemOpen
  \bibfield  {author} {\bibinfo {author} {\bibfnamefont {S.}~\bibnamefont
  {Shalgar}}\ and\ \bibinfo {author} {\bibfnamefont {I.}~\bibnamefont
  {Tamborra}},\ }\href {https://doi.org/10.1103/PhysRevD.107.063025} {\bibfield
   {journal} {\bibinfo  {journal} {Phys. Rev. D}\ }\textbf {\bibinfo {volume}
  {107}},\ \bibinfo {pages} {063025} (\bibinfo {year}
  {2023}{\natexlab{a}})}\BibitemShut {NoStop}%
\bibitem [{\citenamefont {Nagakura}(2023)}]{PhysRevLett.130.211401}%
  \BibitemOpen
  \bibfield  {author} {\bibinfo {author} {\bibfnamefont {H.}~\bibnamefont
  {Nagakura}},\ }\href {https://doi.org/10.1103/PhysRevLett.130.211401}
  {\bibfield  {journal} {\bibinfo  {journal} {Phys. Rev. Lett.}\ }\textbf
  {\bibinfo {volume} {130}},\ \bibinfo {pages} {211401} (\bibinfo {year}
  {2023})}\BibitemShut {NoStop}%
\bibitem [{\citenamefont {Nagakura}\ and\ \citenamefont
  {Zaizen}(2023)}]{nagakura2023basic}%
  \BibitemOpen
  \bibfield  {author} {\bibinfo {author} {\bibfnamefont {H.}~\bibnamefont
  {Nagakura}}\ and\ \bibinfo {author} {\bibfnamefont {M.}~\bibnamefont
  {Zaizen}},\ }\href@noop {} {\bibinfo {title} {Basic characteristics of
  neutrino flavor conversions in the post-shock regions of core-collapse
  supernova}} (\bibinfo {year} {2023}),\ \Eprint
  {https://arxiv.org/abs/2308.14800} {arXiv:2308.14800 [astro-ph.HE]}
  \BibitemShut {NoStop}%
\bibitem [{\citenamefont {Nagakura}\ \emph {et~al.}(2020)\citenamefont
  {Nagakura}, \citenamefont {Burrows}, \citenamefont {Radice},\ and\
  \citenamefont {Vartanyan}}]{10.1093/mnras/staa261}%
  \BibitemOpen
  \bibfield  {author} {\bibinfo {author} {\bibfnamefont {H.}~\bibnamefont
  {Nagakura}}, \bibinfo {author} {\bibfnamefont {A.}~\bibnamefont {Burrows}},
  \bibinfo {author} {\bibfnamefont {D.}~\bibnamefont {Radice}},\ and\ \bibinfo
  {author} {\bibfnamefont {D.}~\bibnamefont {Vartanyan}},\ }\href
  {https://doi.org/10.1093/mnras/staa261} {\bibfield  {journal} {\bibinfo
  {journal} {Monthly Notices of the Royal Astronomical Society}\ }\textbf
  {\bibinfo {volume} {492}},\ \bibinfo {pages} {5764} (\bibinfo {year}
  {2020})},\ \Eprint
  {https://arxiv.org/abs/https://academic.oup.com/mnras/article-pdf/492/4/5764/32432608/staa261.pdf}
  {https://academic.oup.com/mnras/article-pdf/492/4/5764/32432608/staa261.pdf}
  \BibitemShut {NoStop}%
\bibitem [{\citenamefont {Johns}(2023)}]{PhysRevLett.130.191001}%
  \BibitemOpen
  \bibfield  {author} {\bibinfo {author} {\bibfnamefont {L.}~\bibnamefont
  {Johns}},\ }\href {https://doi.org/10.1103/PhysRevLett.130.191001} {\bibfield
   {journal} {\bibinfo  {journal} {Phys. Rev. Lett.}\ }\textbf {\bibinfo
  {volume} {130}},\ \bibinfo {pages} {191001} (\bibinfo {year}
  {2023})}\BibitemShut {NoStop}%
\bibitem [{\citenamefont {Johns}\ and\ \citenamefont
  {Xiong}(2022)}]{PhysRevD.106.103029}%
  \BibitemOpen
  \bibfield  {author} {\bibinfo {author} {\bibfnamefont {L.}~\bibnamefont
  {Johns}}\ and\ \bibinfo {author} {\bibfnamefont {Z.}~\bibnamefont {Xiong}},\
  }\href {https://doi.org/10.1103/PhysRevD.106.103029} {\bibfield  {journal}
  {\bibinfo  {journal} {Phys. Rev. D}\ }\textbf {\bibinfo {volume} {106}},\
  \bibinfo {pages} {103029} (\bibinfo {year} {2022})}\BibitemShut {NoStop}%
\bibitem [{\citenamefont {Lin}\ and\ \citenamefont {Duan}(2022)}]{duan}%
  \BibitemOpen
  \bibfield  {author} {\bibinfo {author} {\bibfnamefont {Y.-C.}\ \bibnamefont
  {Lin}}\ and\ \bibinfo {author} {\bibfnamefont {H.}~\bibnamefont {Duan}},\
  }\href {https://doi.org/10.48550/ARXIV.2210.09218} {\bibinfo {title}
  {Collision-induced flavor instability in dense neutrino gases with
  energy-dependent scattering}} (\bibinfo {year} {2022})\BibitemShut {NoStop}%
\bibitem [{\citenamefont {Xiong}\ \emph
  {et~al.}(2023{\natexlab{b}})\citenamefont {Xiong}, \citenamefont {Wu},
  \citenamefont {Mart\'{\i}nez-Pinedo}, \citenamefont {Fischer}, \citenamefont
  {George}, \citenamefont {Lin},\ and\ \citenamefont
  {Johns}}]{PhysRevD.107.083016}%
  \BibitemOpen
  \bibfield  {author} {\bibinfo {author} {\bibfnamefont {Z.}~\bibnamefont
  {Xiong}}, \bibinfo {author} {\bibfnamefont {M.-R.}\ \bibnamefont {Wu}},
  \bibinfo {author} {\bibfnamefont {G.}~\bibnamefont {Mart\'{\i}nez-Pinedo}},
  \bibinfo {author} {\bibfnamefont {T.}~\bibnamefont {Fischer}}, \bibinfo
  {author} {\bibfnamefont {M.}~\bibnamefont {George}}, \bibinfo {author}
  {\bibfnamefont {C.-Y.}\ \bibnamefont {Lin}},\ and\ \bibinfo {author}
  {\bibfnamefont {L.}~\bibnamefont {Johns}},\ }\href
  {https://doi.org/10.1103/PhysRevD.107.083016} {\bibfield  {journal} {\bibinfo
   {journal} {Phys. Rev. D}\ }\textbf {\bibinfo {volume} {107}},\ \bibinfo
  {pages} {083016} (\bibinfo {year} {2023}{\natexlab{b}})}\BibitemShut
  {NoStop}%
\bibitem [{\citenamefont {Xiong}\ \emph {et~al.}(2022)\citenamefont {Xiong},
  \citenamefont {Johns}, \citenamefont {Wu},\ and\ \citenamefont
  {Duan}}]{2022}%
  \BibitemOpen
  \bibfield  {author} {\bibinfo {author} {\bibfnamefont {Z.}~\bibnamefont
  {Xiong}}, \bibinfo {author} {\bibfnamefont {L.}~\bibnamefont {Johns}},
  \bibinfo {author} {\bibfnamefont {M.-R.}\ \bibnamefont {Wu}},\ and\ \bibinfo
  {author} {\bibfnamefont {H.}~\bibnamefont {Duan}},\ }\href
  {https://doi.org/10.48550/ARXIV.2212.03750} {\bibinfo {title} {Collisional
  flavor instability in dense neutrino gases}} (\bibinfo {year}
  {2022})\BibitemShut {NoStop}%
\bibitem [{\citenamefont {Liu}\ \emph {et~al.}(2023)\citenamefont {Liu},
  \citenamefont {Zaizen},\ and\ \citenamefont {Yamada}}]{PhysRevD.107.123011}%
  \BibitemOpen
  \bibfield  {author} {\bibinfo {author} {\bibfnamefont {J.}~\bibnamefont
  {Liu}}, \bibinfo {author} {\bibfnamefont {M.}~\bibnamefont {Zaizen}},\ and\
  \bibinfo {author} {\bibfnamefont {S.}~\bibnamefont {Yamada}},\ }\href
  {https://doi.org/10.1103/PhysRevD.107.123011} {\bibfield  {journal} {\bibinfo
   {journal} {Phys. Rev. D}\ }\textbf {\bibinfo {volume} {107}},\ \bibinfo
  {pages} {123011} (\bibinfo {year} {2023})}\BibitemShut {NoStop}%
\bibitem [{\citenamefont {Kato}\ \emph {et~al.}(2023)\citenamefont {Kato},
  \citenamefont {Nagakura},\ and\ \citenamefont {Johns}}]{kato2023collisional}%
  \BibitemOpen
  \bibfield  {author} {\bibinfo {author} {\bibfnamefont {C.}~\bibnamefont
  {Kato}}, \bibinfo {author} {\bibfnamefont {H.}~\bibnamefont {Nagakura}},\
  and\ \bibinfo {author} {\bibfnamefont {L.}~\bibnamefont {Johns}},\
  }\href@noop {} {\bibinfo {title} {Collisional flavor swap with neutrino
  self-interactions}} (\bibinfo {year} {2023}),\ \Eprint
  {https://arxiv.org/abs/2309.02619} {arXiv:2309.02619 [astro-ph.HE]}
  \BibitemShut {NoStop}%
\bibitem [{\citenamefont {Shalgar}\ and\ \citenamefont
  {Tamborra}(2023{\natexlab{b}})}]{shalgar2023neutrinos}%
  \BibitemOpen
  \bibfield  {author} {\bibinfo {author} {\bibfnamefont {S.}~\bibnamefont
  {Shalgar}}\ and\ \bibinfo {author} {\bibfnamefont {I.}~\bibnamefont
  {Tamborra}},\ }\href@noop {} {\bibinfo {title} {Do neutrinos become flavor
  unstable due to collisions with matter in the supernova decoupling region?}}
  (\bibinfo {year} {2023}{\natexlab{b}}),\ \Eprint
  {https://arxiv.org/abs/2307.10366} {arXiv:2307.10366 [astro-ph.HE]}
  \BibitemShut {NoStop}%
\bibitem [{mpa()}]{mpaarchive}%
  \BibitemOpen
  \href@noop {} {\bibinfo {title} {The garching core-collapse supernova
  archive}},\ \bibinfo {howpublished}
  {\url{https://wwwmpa.mpa-garching.mpg.de/ccsnarchive},
  \url{https://wwwmpa.mpa-garching.mpg.de/ccsnarchive/data/Bollig2016/}}\BibitemShut
  {NoStop}%
\bibitem [{\citenamefont {{Mezzacappa}}\ and\ \citenamefont
  {{Bruenn}}(1993)}]{1993ApJ...405..669M}%
  \BibitemOpen
  \bibfield  {author} {\bibinfo {author} {\bibfnamefont {A.}~\bibnamefont
  {{Mezzacappa}}}\ and\ \bibinfo {author} {\bibfnamefont {S.~W.}\ \bibnamefont
  {{Bruenn}}},\ }\href {https://doi.org/10.1086/172395} {\bibfield  {journal}
  {\bibinfo  {journal} {\apj}\ }\textbf {\bibinfo {volume} {405}},\ \bibinfo
  {pages} {669} (\bibinfo {year} {1993})}\BibitemShut {NoStop}%
\bibitem [{\citenamefont {Richers}\ \emph {et~al.}(2019)\citenamefont
  {Richers}, \citenamefont {McLaughlin}, \citenamefont {Kneller},\ and\
  \citenamefont {Vlasenko}}]{PhysRevD.99.123014}%
  \BibitemOpen
  \bibfield  {author} {\bibinfo {author} {\bibfnamefont {S.~A.}\ \bibnamefont
  {Richers}}, \bibinfo {author} {\bibfnamefont {G.~C.}\ \bibnamefont
  {McLaughlin}}, \bibinfo {author} {\bibfnamefont {J.~P.}\ \bibnamefont
  {Kneller}},\ and\ \bibinfo {author} {\bibfnamefont {A.}~\bibnamefont
  {Vlasenko}},\ }\href {https://doi.org/10.1103/PhysRevD.99.123014} {\bibfield
  {journal} {\bibinfo  {journal} {Phys. Rev. D}\ }\textbf {\bibinfo {volume}
  {99}},\ \bibinfo {pages} {123014} (\bibinfo {year} {2019})}\BibitemShut
  {NoStop}%
\bibitem [{\citenamefont {Nagakura}\ \emph {et~al.}(2014)\citenamefont
  {Nagakura}, \citenamefont {Sumiyoshi},\ and\ \citenamefont
  {Yamada}}]{Nagakura_2014}%
  \BibitemOpen
  \bibfield  {author} {\bibinfo {author} {\bibfnamefont {H.}~\bibnamefont
  {Nagakura}}, \bibinfo {author} {\bibfnamefont {K.}~\bibnamefont
  {Sumiyoshi}},\ and\ \bibinfo {author} {\bibfnamefont {S.}~\bibnamefont
  {Yamada}},\ }\href {https://doi.org/10.1088/0067-0049/214/2/16} {\bibfield
  {journal} {\bibinfo  {journal} {The Astrophysical Journal Supplement Series}\
  }\textbf {\bibinfo {volume} {214}},\ \bibinfo {pages} {16} (\bibinfo {year}
  {2014})}\BibitemShut {NoStop}%
\bibitem [{\citenamefont {Nagakura}\ \emph {et~al.}(2017)\citenamefont
  {Nagakura}, \citenamefont {Iwakami}, \citenamefont {Furusawa}, \citenamefont
  {Sumiyoshi}, \citenamefont {Yamada}, \citenamefont {Matsufuru},\ and\
  \citenamefont {Imakura}}]{Nagakura_2017}%
  \BibitemOpen
  \bibfield  {author} {\bibinfo {author} {\bibfnamefont {H.}~\bibnamefont
  {Nagakura}}, \bibinfo {author} {\bibfnamefont {W.}~\bibnamefont {Iwakami}},
  \bibinfo {author} {\bibfnamefont {S.}~\bibnamefont {Furusawa}}, \bibinfo
  {author} {\bibfnamefont {K.}~\bibnamefont {Sumiyoshi}}, \bibinfo {author}
  {\bibfnamefont {S.}~\bibnamefont {Yamada}}, \bibinfo {author} {\bibfnamefont
  {H.}~\bibnamefont {Matsufuru}},\ and\ \bibinfo {author} {\bibfnamefont
  {A.}~\bibnamefont {Imakura}},\ }\href
  {https://doi.org/10.3847/1538-4365/aa69ea} {\bibfield  {journal} {\bibinfo
  {journal} {The Astrophysical Journal Supplement Series}\ }\textbf {\bibinfo
  {volume} {229}},\ \bibinfo {pages} {42} (\bibinfo {year} {2017})}\BibitemShut
  {NoStop}%
\bibitem [{\citenamefont {Nagakura}\ \emph
  {et~al.}(2019{\natexlab{b}})\citenamefont {Nagakura}, \citenamefont
  {Sumiyoshi},\ and\ \citenamefont {Yamada}}]{Nagakura_2019c}%
  \BibitemOpen
  \bibfield  {author} {\bibinfo {author} {\bibfnamefont {H.}~\bibnamefont
  {Nagakura}}, \bibinfo {author} {\bibfnamefont {K.}~\bibnamefont
  {Sumiyoshi}},\ and\ \bibinfo {author} {\bibfnamefont {S.}~\bibnamefont
  {Yamada}},\ }\href {https://doi.org/10.3847/1538-4357/ab2189} {\bibfield
  {journal} {\bibinfo  {journal} {The Astrophysical Journal}\ }\textbf
  {\bibinfo {volume} {878}},\ \bibinfo {pages} {160} (\bibinfo {year}
  {2019}{\natexlab{b}})}\BibitemShut {NoStop}%
\bibitem [{\citenamefont {Nagakura}\ \emph {et~al.}(2018)\citenamefont
  {Nagakura}, \citenamefont {Iwakami}, \citenamefont {Furusawa}, \citenamefont
  {Okawa}, \citenamefont {Harada}, \citenamefont {Sumiyoshi}, \citenamefont
  {Yamada}, \citenamefont {Matsufuru},\ and\ \citenamefont
  {Imakura}}]{Nagakura_2018}%
  \BibitemOpen
  \bibfield  {author} {\bibinfo {author} {\bibfnamefont {H.}~\bibnamefont
  {Nagakura}}, \bibinfo {author} {\bibfnamefont {W.}~\bibnamefont {Iwakami}},
  \bibinfo {author} {\bibfnamefont {S.}~\bibnamefont {Furusawa}}, \bibinfo
  {author} {\bibfnamefont {H.}~\bibnamefont {Okawa}}, \bibinfo {author}
  {\bibfnamefont {A.}~\bibnamefont {Harada}}, \bibinfo {author} {\bibfnamefont
  {K.}~\bibnamefont {Sumiyoshi}}, \bibinfo {author} {\bibfnamefont
  {S.}~\bibnamefont {Yamada}}, \bibinfo {author} {\bibfnamefont
  {H.}~\bibnamefont {Matsufuru}},\ and\ \bibinfo {author} {\bibfnamefont
  {A.}~\bibnamefont {Imakura}},\ }\href
  {https://doi.org/10.3847/1538-4357/aaac29} {\bibfield  {journal} {\bibinfo
  {journal} {The Astrophysical Journal}\ }\textbf {\bibinfo {volume} {854}},\
  \bibinfo {pages} {136} (\bibinfo {year} {2018})}\BibitemShut {NoStop}%
\bibitem [{\citenamefont {Harada}\ \emph {et~al.}(2019)\citenamefont {Harada},
  \citenamefont {Nagakura}, \citenamefont {Iwakami}, \citenamefont {Okawa},
  \citenamefont {Furusawa}, \citenamefont {Matsufuru}, \citenamefont
  {Sumiyoshi},\ and\ \citenamefont {Yamada}}]{Harada_2019}%
  \BibitemOpen
  \bibfield  {author} {\bibinfo {author} {\bibfnamefont {A.}~\bibnamefont
  {Harada}}, \bibinfo {author} {\bibfnamefont {H.}~\bibnamefont {Nagakura}},
  \bibinfo {author} {\bibfnamefont {W.}~\bibnamefont {Iwakami}}, \bibinfo
  {author} {\bibfnamefont {H.}~\bibnamefont {Okawa}}, \bibinfo {author}
  {\bibfnamefont {S.}~\bibnamefont {Furusawa}}, \bibinfo {author}
  {\bibfnamefont {H.}~\bibnamefont {Matsufuru}}, \bibinfo {author}
  {\bibfnamefont {K.}~\bibnamefont {Sumiyoshi}},\ and\ \bibinfo {author}
  {\bibfnamefont {S.}~\bibnamefont {Yamada}},\ }\href
  {https://doi.org/10.3847/1538-4357/ab0203} {\bibfield  {journal} {\bibinfo
  {journal} {The Astrophysical Journal}\ }\textbf {\bibinfo {volume} {872}},\
  \bibinfo {pages} {181} (\bibinfo {year} {2019})}\BibitemShut {NoStop}%
\bibitem [{\citenamefont {Nagakura}\ \emph
  {et~al.}(2019{\natexlab{c}})\citenamefont {Nagakura}, \citenamefont
  {Sumiyoshi},\ and\ \citenamefont {Yamada}}]{Nagakura_2019acceleration}%
  \BibitemOpen
  \bibfield  {author} {\bibinfo {author} {\bibfnamefont {H.}~\bibnamefont
  {Nagakura}}, \bibinfo {author} {\bibfnamefont {K.}~\bibnamefont
  {Sumiyoshi}},\ and\ \bibinfo {author} {\bibfnamefont {S.}~\bibnamefont
  {Yamada}},\ }\href {https://doi.org/10.3847/2041-8213/ab30ca} {\bibfield
  {journal} {\bibinfo  {journal} {The Astrophysical Journal Letters}\ }\textbf
  {\bibinfo {volume} {880}},\ \bibinfo {pages} {L28} (\bibinfo {year}
  {2019}{\natexlab{c}})}\BibitemShut {NoStop}%
\bibitem [{\citenamefont {Iwakami}\ \emph {et~al.}(2020)\citenamefont
  {Iwakami}, \citenamefont {Okawa}, \citenamefont {Nagakura}, \citenamefont
  {Harada}, \citenamefont {Furusawa}, \citenamefont {Sumiyoshi}, \citenamefont
  {Matsufuru},\ and\ \citenamefont {Yamada}}]{Iwakami_2020}%
  \BibitemOpen
  \bibfield  {author} {\bibinfo {author} {\bibfnamefont {W.}~\bibnamefont
  {Iwakami}}, \bibinfo {author} {\bibfnamefont {H.}~\bibnamefont {Okawa}},
  \bibinfo {author} {\bibfnamefont {H.}~\bibnamefont {Nagakura}}, \bibinfo
  {author} {\bibfnamefont {A.}~\bibnamefont {Harada}}, \bibinfo {author}
  {\bibfnamefont {S.}~\bibnamefont {Furusawa}}, \bibinfo {author}
  {\bibfnamefont {K.}~\bibnamefont {Sumiyoshi}}, \bibinfo {author}
  {\bibfnamefont {H.}~\bibnamefont {Matsufuru}},\ and\ \bibinfo {author}
  {\bibfnamefont {S.}~\bibnamefont {Yamada}},\ }\href
  {https://doi.org/10.3847/1538-4357/abb8cf} {\bibfield  {journal} {\bibinfo
  {journal} {The Astrophysical Journal}\ }\textbf {\bibinfo {volume} {903}},\
  \bibinfo {pages} {82} (\bibinfo {year} {2020})}\BibitemShut {NoStop}%
\bibitem [{\citenamefont {Iwakami}\ \emph {et~al.}(2022)\citenamefont
  {Iwakami}, \citenamefont {Harada}, \citenamefont {Nagakura}, \citenamefont
  {Akaho}, \citenamefont {Okawa}, \citenamefont {Furusawa}, \citenamefont
  {Matsufuru}, \citenamefont {Sumiyoshi},\ and\ \citenamefont
  {Yamada}}]{Iwakami_2022}%
  \BibitemOpen
  \bibfield  {author} {\bibinfo {author} {\bibfnamefont {W.}~\bibnamefont
  {Iwakami}}, \bibinfo {author} {\bibfnamefont {A.}~\bibnamefont {Harada}},
  \bibinfo {author} {\bibfnamefont {H.}~\bibnamefont {Nagakura}}, \bibinfo
  {author} {\bibfnamefont {R.}~\bibnamefont {Akaho}}, \bibinfo {author}
  {\bibfnamefont {H.}~\bibnamefont {Okawa}}, \bibinfo {author} {\bibfnamefont
  {S.}~\bibnamefont {Furusawa}}, \bibinfo {author} {\bibfnamefont
  {H.}~\bibnamefont {Matsufuru}}, \bibinfo {author} {\bibfnamefont
  {K.}~\bibnamefont {Sumiyoshi}},\ and\ \bibinfo {author} {\bibfnamefont
  {S.}~\bibnamefont {Yamada}},\ }\href
  {https://doi.org/10.3847/1538-4357/ac714b} {\bibfield  {journal} {\bibinfo
  {journal} {The Astrophysical Journal}\ }\textbf {\bibinfo {volume} {933}},\
  \bibinfo {pages} {91} (\bibinfo {year} {2022})}\BibitemShut {NoStop}%
\bibitem [{\citenamefont {Togashi}\ and\ \citenamefont
  {Takano}(2013)}]{TOGASHI201353}%
  \BibitemOpen
  \bibfield  {author} {\bibinfo {author} {\bibfnamefont {H.}~\bibnamefont
  {Togashi}}\ and\ \bibinfo {author} {\bibfnamefont {M.}~\bibnamefont
  {Takano}},\ }\href
  {https://doi.org/https://doi.org/10.1016/j.nuclphysa.2013.02.014} {\bibfield
  {journal} {\bibinfo  {journal} {Nuclear Physics A}\ }\textbf {\bibinfo
  {volume} {902}},\ \bibinfo {pages} {53} (\bibinfo {year} {2013})}\BibitemShut
  {NoStop}%
\bibitem [{\citenamefont {Togashi}\ \emph {et~al.}(2017)\citenamefont
  {Togashi}, \citenamefont {Nakazato}, \citenamefont {Takehara}, \citenamefont
  {Yamamuro}, \citenamefont {Suzuki},\ and\ \citenamefont
  {Takano}}]{TOGASHI201778}%
  \BibitemOpen
  \bibfield  {author} {\bibinfo {author} {\bibfnamefont {H.}~\bibnamefont
  {Togashi}}, \bibinfo {author} {\bibfnamefont {K.}~\bibnamefont {Nakazato}},
  \bibinfo {author} {\bibfnamefont {Y.}~\bibnamefont {Takehara}}, \bibinfo
  {author} {\bibfnamefont {S.}~\bibnamefont {Yamamuro}}, \bibinfo {author}
  {\bibfnamefont {H.}~\bibnamefont {Suzuki}},\ and\ \bibinfo {author}
  {\bibfnamefont {M.}~\bibnamefont {Takano}},\ }\href
  {https://doi.org/https://doi.org/10.1016/j.nuclphysa.2017.02.010} {\bibfield
  {journal} {\bibinfo  {journal} {Nuclear Physics A}\ }\textbf {\bibinfo
  {volume} {961}},\ \bibinfo {pages} {78} (\bibinfo {year} {2017})}\BibitemShut
  {NoStop}%
\bibitem [{\citenamefont {Furusawa}\ \emph {et~al.}(2017)\citenamefont
  {Furusawa}, \citenamefont {Togashi}, \citenamefont {Nagakura}, \citenamefont
  {Sumiyoshi}, \citenamefont {Yamada}, \citenamefont {Suzuki},\ and\
  \citenamefont {Takano}}]{Furusawa_2017}%
  \BibitemOpen
  \bibfield  {author} {\bibinfo {author} {\bibfnamefont {S.}~\bibnamefont
  {Furusawa}}, \bibinfo {author} {\bibfnamefont {H.}~\bibnamefont {Togashi}},
  \bibinfo {author} {\bibfnamefont {H.}~\bibnamefont {Nagakura}}, \bibinfo
  {author} {\bibfnamefont {K.}~\bibnamefont {Sumiyoshi}}, \bibinfo {author}
  {\bibfnamefont {S.}~\bibnamefont {Yamada}}, \bibinfo {author} {\bibfnamefont
  {H.}~\bibnamefont {Suzuki}},\ and\ \bibinfo {author} {\bibfnamefont
  {M.}~\bibnamefont {Takano}},\ }\href
  {https://doi.org/10.1088/1361-6471/aa7f35} {\bibfield  {journal} {\bibinfo
  {journal} {Journal of Physics G: Nuclear and Particle Physics}\ }\textbf
  {\bibinfo {volume} {44}},\ \bibinfo {pages} {094001} (\bibinfo {year}
  {2017})}\BibitemShut {NoStop}%
\bibitem [{\citenamefont {Nagakura}\ \emph
  {et~al.}(2019{\natexlab{d}})\citenamefont {Nagakura}, \citenamefont
  {Furusawa}, \citenamefont {Togashi}, \citenamefont {Richers}, \citenamefont
  {Sumiyoshi},\ and\ \citenamefont {Yamada}}]{Nagakura_2019a}%
  \BibitemOpen
  \bibfield  {author} {\bibinfo {author} {\bibfnamefont {H.}~\bibnamefont
  {Nagakura}}, \bibinfo {author} {\bibfnamefont {S.}~\bibnamefont {Furusawa}},
  \bibinfo {author} {\bibfnamefont {H.}~\bibnamefont {Togashi}}, \bibinfo
  {author} {\bibfnamefont {S.}~\bibnamefont {Richers}}, \bibinfo {author}
  {\bibfnamefont {K.}~\bibnamefont {Sumiyoshi}},\ and\ \bibinfo {author}
  {\bibfnamefont {S.}~\bibnamefont {Yamada}},\ }\href
  {https://doi.org/10.3847/1538-4365/aafac9} {\bibfield  {journal} {\bibinfo
  {journal} {The Astrophysical Journal Supplement Series}\ }\textbf {\bibinfo
  {volume} {240}},\ \bibinfo {pages} {38} (\bibinfo {year}
  {2019}{\natexlab{d}})}\BibitemShut {NoStop}%
\bibitem [{\citenamefont {Sumiyoshi}\ and\ \citenamefont
  {Yamada}(2012)}]{Sumiyoshi_2012}%
  \BibitemOpen
  \bibfield  {author} {\bibinfo {author} {\bibfnamefont {K.}~\bibnamefont
  {Sumiyoshi}}\ and\ \bibinfo {author} {\bibfnamefont {S.}~\bibnamefont
  {Yamada}},\ }\href {https://doi.org/10.1088/0067-0049/199/1/17} {\bibfield
  {journal} {\bibinfo  {journal} {The Astrophysical Journal Supplement Series}\
  }\textbf {\bibinfo {volume} {199}},\ \bibinfo {pages} {17} (\bibinfo {year}
  {2012})}\BibitemShut {NoStop}%
\bibitem [{\citenamefont {Woosley}\ \emph {et~al.}(2002)\citenamefont
  {Woosley}, \citenamefont {Heger},\ and\ \citenamefont
  {Weaver}}]{RevModPhys.74.1015}%
  \BibitemOpen
  \bibfield  {author} {\bibinfo {author} {\bibfnamefont {S.~E.}\ \bibnamefont
  {Woosley}}, \bibinfo {author} {\bibfnamefont {A.}~\bibnamefont {Heger}},\
  and\ \bibinfo {author} {\bibfnamefont {T.~A.}\ \bibnamefont {Weaver}},\
  }\href {https://doi.org/10.1103/RevModPhys.74.1015} {\bibfield  {journal}
  {\bibinfo  {journal} {Rev. Mod. Phys.}\ }\textbf {\bibinfo {volume} {74}},\
  \bibinfo {pages} {1015} (\bibinfo {year} {2002})}\BibitemShut {NoStop}%
\bibitem [{\citenamefont {Burrows}\ \emph {et~al.}(2023)\citenamefont
  {Burrows}, \citenamefont {Vartanyan},\ and\ \citenamefont
  {Wang}}]{burrows2023blackhole}%
  \BibitemOpen
  \bibfield  {author} {\bibinfo {author} {\bibfnamefont {A.}~\bibnamefont
  {Burrows}}, \bibinfo {author} {\bibfnamefont {D.}~\bibnamefont {Vartanyan}},\
  and\ \bibinfo {author} {\bibfnamefont {T.}~\bibnamefont {Wang}},\ }\href@noop
  {} {\bibinfo {title} {Black-hole formation accompanied by the supernova
  explosion of a 40-m$_{\odot}$ progenitor star}} (\bibinfo {year} {2023}),\
  \Eprint {https://arxiv.org/abs/2308.05798} {arXiv:2308.05798 [astro-ph.SR]}
  \BibitemShut {NoStop}%
\bibitem [{\citenamefont {Esteban-Pretel}\ \emph {et~al.}(2008)\citenamefont
  {Esteban-Pretel}, \citenamefont {Mirizzi}, \citenamefont {Pastor},
  \citenamefont {Tom\`as}, \citenamefont {Raffelt}, \citenamefont {Serpico},\
  and\ \citenamefont {Sigl}}]{PhysRevD.78.085012}%
  \BibitemOpen
  \bibfield  {author} {\bibinfo {author} {\bibfnamefont {A.}~\bibnamefont
  {Esteban-Pretel}}, \bibinfo {author} {\bibfnamefont {A.}~\bibnamefont
  {Mirizzi}}, \bibinfo {author} {\bibfnamefont {S.}~\bibnamefont {Pastor}},
  \bibinfo {author} {\bibfnamefont {R.}~\bibnamefont {Tom\`as}}, \bibinfo
  {author} {\bibfnamefont {G.~G.}\ \bibnamefont {Raffelt}}, \bibinfo {author}
  {\bibfnamefont {P.~D.}\ \bibnamefont {Serpico}},\ and\ \bibinfo {author}
  {\bibfnamefont {G.}~\bibnamefont {Sigl}},\ }\href
  {https://doi.org/10.1103/PhysRevD.78.085012} {\bibfield  {journal} {\bibinfo
  {journal} {Phys. Rev. D}\ }\textbf {\bibinfo {volume} {78}},\ \bibinfo
  {pages} {085012} (\bibinfo {year} {2008})}\BibitemShut {NoStop}%
\bibitem [{\citenamefont {Dasgupta}\ \emph {et~al.}(2012)\citenamefont
  {Dasgupta}, \citenamefont {O'Connor},\ and\ \citenamefont
  {Ott}}]{PhysRevD.85.065008}%
  \BibitemOpen
  \bibfield  {author} {\bibinfo {author} {\bibfnamefont {B.}~\bibnamefont
  {Dasgupta}}, \bibinfo {author} {\bibfnamefont {E.~P.}\ \bibnamefont
  {O'Connor}},\ and\ \bibinfo {author} {\bibfnamefont {C.~D.}\ \bibnamefont
  {Ott}},\ }\href {https://doi.org/10.1103/PhysRevD.85.065008} {\bibfield
  {journal} {\bibinfo  {journal} {Phys. Rev. D}\ }\textbf {\bibinfo {volume}
  {85}},\ \bibinfo {pages} {065008} (\bibinfo {year} {2012})}\BibitemShut
  {NoStop}%
\bibitem [{\citenamefont {Chakraborty}\ \emph {et~al.}(2011)\citenamefont
  {Chakraborty}, \citenamefont {Fischer}, \citenamefont {Mirizzi},
  \citenamefont {Saviano},\ and\ \citenamefont
  {Tom\`as}}]{PhysRevLett.107.151101}%
  \BibitemOpen
  \bibfield  {author} {\bibinfo {author} {\bibfnamefont {S.}~\bibnamefont
  {Chakraborty}}, \bibinfo {author} {\bibfnamefont {T.}~\bibnamefont
  {Fischer}}, \bibinfo {author} {\bibfnamefont {A.}~\bibnamefont {Mirizzi}},
  \bibinfo {author} {\bibfnamefont {N.}~\bibnamefont {Saviano}},\ and\ \bibinfo
  {author} {\bibfnamefont {R.}~\bibnamefont {Tom\`as}},\ }\href
  {https://doi.org/10.1103/PhysRevLett.107.151101} {\bibfield  {journal}
  {\bibinfo  {journal} {Phys. Rev. Lett.}\ }\textbf {\bibinfo {volume} {107}},\
  \bibinfo {pages} {151101} (\bibinfo {year} {2011})}\BibitemShut {NoStop}%
\bibitem [{\citenamefont {Sarikas}\ \emph {et~al.}(2012)\citenamefont
  {Sarikas}, \citenamefont {Tamborra}, \citenamefont {Raffelt}, \citenamefont
  {H\"udepohl},\ and\ \citenamefont {Janka}}]{PhysRevD.85.113007}%
  \BibitemOpen
  \bibfield  {author} {\bibinfo {author} {\bibfnamefont {S.}~\bibnamefont
  {Sarikas}}, \bibinfo {author} {\bibfnamefont {I.}~\bibnamefont {Tamborra}},
  \bibinfo {author} {\bibfnamefont {G.}~\bibnamefont {Raffelt}}, \bibinfo
  {author} {\bibfnamefont {L.}~\bibnamefont {H\"udepohl}},\ and\ \bibinfo
  {author} {\bibfnamefont {H.-T.}\ \bibnamefont {Janka}},\ }\href
  {https://doi.org/10.1103/PhysRevD.85.113007} {\bibfield  {journal} {\bibinfo
  {journal} {Phys. Rev. D}\ }\textbf {\bibinfo {volume} {85}},\ \bibinfo
  {pages} {113007} (\bibinfo {year} {2012})}\BibitemShut {NoStop}%
\bibitem [{\citenamefont {Sugiura}\ \emph {et~al.}(2022)\citenamefont
  {Sugiura}, \citenamefont {Furusawa}, \citenamefont {Sumiyoshi},\ and\
  \citenamefont {Yamada}}]{10.1093/ptep/ptac118}%
  \BibitemOpen
  \bibfield  {author} {\bibinfo {author} {\bibfnamefont {K.}~\bibnamefont
  {Sugiura}}, \bibinfo {author} {\bibfnamefont {S.}~\bibnamefont {Furusawa}},
  \bibinfo {author} {\bibfnamefont {K.}~\bibnamefont {Sumiyoshi}},\ and\
  \bibinfo {author} {\bibfnamefont {S.}~\bibnamefont {Yamada}},\ }\href
  {https://doi.org/10.1093/ptep/ptac118} {\bibfield  {journal} {\bibinfo
  {journal} {Progress of Theoretical and Experimental Physics}\ }\textbf
  {\bibinfo {volume} {2022}},\ \bibinfo {pages} {113E01} (\bibinfo {year}
  {2022})},\ \Eprint
  {https://arxiv.org/abs/https://academic.oup.com/ptep/article-pdf/2022/11/113E01/46781886/ptac118.pdf}
  {https://academic.oup.com/ptep/article-pdf/2022/11/113E01/46781886/ptac118.pdf}
  \BibitemShut {NoStop}%
\bibitem [{\citenamefont {Bollig}\ \emph {et~al.}(2017)\citenamefont {Bollig},
  \citenamefont {Janka}, \citenamefont {Lohs}, \citenamefont
  {Mart\'{\i}nez-Pinedo}, \citenamefont {Horowitz},\ and\ \citenamefont
  {Melson}}]{PhysRevLett.119.242702}%
  \BibitemOpen
  \bibfield  {author} {\bibinfo {author} {\bibfnamefont {R.}~\bibnamefont
  {Bollig}}, \bibinfo {author} {\bibfnamefont {H.-T.}\ \bibnamefont {Janka}},
  \bibinfo {author} {\bibfnamefont {A.}~\bibnamefont {Lohs}}, \bibinfo {author}
  {\bibfnamefont {G.}~\bibnamefont {Mart\'{\i}nez-Pinedo}}, \bibinfo {author}
  {\bibfnamefont {C.~J.}\ \bibnamefont {Horowitz}},\ and\ \bibinfo {author}
  {\bibfnamefont {T.}~\bibnamefont {Melson}},\ }\href
  {https://doi.org/10.1103/PhysRevLett.119.242702} {\bibfield  {journal}
  {\bibinfo  {journal} {Phys. Rev. Lett.}\ }\textbf {\bibinfo {volume} {119}},\
  \bibinfo {pages} {242702} (\bibinfo {year} {2017})}\BibitemShut {NoStop}%
\bibitem [{\citenamefont {Guo}\ \emph {et~al.}(2020)\citenamefont {Guo},
  \citenamefont {Mart\'{\i}nez-Pinedo}, \citenamefont {Lohs},\ and\
  \citenamefont {Fischer}}]{PhysRevD.102.023037}%
  \BibitemOpen
  \bibfield  {author} {\bibinfo {author} {\bibfnamefont {G.}~\bibnamefont
  {Guo}}, \bibinfo {author} {\bibfnamefont {G.}~\bibnamefont
  {Mart\'{\i}nez-Pinedo}}, \bibinfo {author} {\bibfnamefont {A.}~\bibnamefont
  {Lohs}},\ and\ \bibinfo {author} {\bibfnamefont {T.}~\bibnamefont
  {Fischer}},\ }\href {https://doi.org/10.1103/PhysRevD.102.023037} {\bibfield
  {journal} {\bibinfo  {journal} {Phys. Rev. D}\ }\textbf {\bibinfo {volume}
  {102}},\ \bibinfo {pages} {023037} (\bibinfo {year} {2020})}\BibitemShut
  {NoStop}%
\bibitem [{\citenamefont {Fischer}\ \emph {et~al.}(2020)\citenamefont
  {Fischer}, \citenamefont {Guo}, \citenamefont {Mart\'{\i}nez-Pinedo},
  \citenamefont {Liebend\"orfer},\ and\ \citenamefont
  {Mezzacappa}}]{PhysRevD.102.123001}%
  \BibitemOpen
  \bibfield  {author} {\bibinfo {author} {\bibfnamefont {T.}~\bibnamefont
  {Fischer}}, \bibinfo {author} {\bibfnamefont {G.}~\bibnamefont {Guo}},
  \bibinfo {author} {\bibfnamefont {G.}~\bibnamefont {Mart\'{\i}nez-Pinedo}},
  \bibinfo {author} {\bibfnamefont {M.}~\bibnamefont {Liebend\"orfer}},\ and\
  \bibinfo {author} {\bibfnamefont {A.}~\bibnamefont {Mezzacappa}},\ }\href
  {https://doi.org/10.1103/PhysRevD.102.123001} {\bibfield  {journal} {\bibinfo
   {journal} {Phys. Rev. D}\ }\textbf {\bibinfo {volume} {102}},\ \bibinfo
  {pages} {123001} (\bibinfo {year} {2020})}\BibitemShut {NoStop}%
\end{thebibliography}%
\bibliographystyle{apsrev4-2}
\end{document}